\documentclass[12pt,preprint]{aastex}
\usepackage{epsf}
\newcommand{\simgt}{\lower.5ex\hbox{$\; \buildrel > \over \sim \;$}}
\newcommand{\simlt}{\lower.5ex\hbox{$\; \buildrel < \over \sim \;$}}
\newcommand{\probI}{P_{\scriptscriptstyle I}}
\newcommand{\bm}[1]{{\mbox{\boldmath${#1}$}}}
\newcommand{\Lbox}{L_{\rm\scriptscriptstyle BOX}}
\slugcomment{RESCEU-96/03\quad UTAP-465/2003}
\shortauthors{Ohta,Kayo\& Taruya}
\shorttitle{Cosmological Density Distribution From Ellipsoidal Collapse}
\begin{document}
%%%%%%%%%%%%%%%%%%%%%%%%%%%%%%%%%%%%%%%%%%%%%%%%%%%%%%%%%%%%%%%%%%%%%%
%%%%%%%%%%%%%%%%%%%%%%%%%%%%%%%%%%%%%%%%%%%%%%%%%%%%%%%%%%%%%%%%%%%%%%
\title{\Large Cosmological Density Distribution Function from the
Ellipsoidal Collapse Model in Real Space}
%%%%%%%%%%%%%%%%%%%%%%%%%%%%%%%%%%%%%%%%%%%%%%%%%%%%%%%%%%%%%%%%%%%%%%
%%%%%%%%%%%%%%%%%%%%%%%%%%%%%%%%%%%%%%%%%%%%%%%%%%%%%%%%%%%%%%%%%%%%%%
%
%
%%%%%%%%%%%%%%%%%%%%%%%%%%%%%%%%%%%%%%%%%%%%%%%%%%%%%%%%%%%%%%%%%%%%%%
\author{Yasuhiro Ohta,\altaffilmark{1} Issha Kayo,\altaffilmark{1} and,
Atsushi Taruya\altaffilmark{1,2}}
\altaffiltext{1}{Department of Physics, School of Science, University of
Tokyo, Tokyo 113, Japan.}
\altaffiltext{2}{Research Center for the Early Universe(RESCEU), School
of Science, University of Tokyo, Tokyo 113, Japan.}
\email{ohta@utap.phys.s.u-tokyo.ac.jp, kayo@utap.phys.s.u-tokyo.ac.jp,
ataruya@utap.phys.s.u-tokyo.ac.jp}
%%%%%%%%%%%%%%%%%%%%%%%%%%%%%%%%%%%%%%%%%%%%%%%%%%%%%%%%%%%%%%%%%%%%%%
%
%
%
%%%%%%%%%%%%%%%%%%%%%%%%%%%%%%%%%%%%%%%%%%%%%%%%%%%%%%%%%%%%%%%%%%%%%%
\begin{abstract}
 We calculate the one-point probability distribution function (PDF) for
 cosmic density $\delta$ in non-linear regime of the gravitational
 evolution. Under the local approximation that the evolution of cosmic
 fluid fields can be characterized by the Lagrangian local dynamics with
 finite degrees of freedom, the analytic expressions of PDF are derived
 taking account of the smoothing effect. The validity and the
 usefulness of the local approximation are then discussed comparing
 those results with N-body simulations in a Gaussian initial
 condition. Adopting the ellipsoidal collapse model (ECM) and the spherical 
 collapse model (SCM) as Lagrangian local dynamics, we found that the 
 PDFs from the local approximation excellently match the simulation 
 results in the case of the cold dark matter initial spectrum. 
 As for the scale-free initial spectra given by $P(k)\propto k^n$, 
 N-body result suffers from spurious numerical effects, which prevent 
 us to give a detailed comparison. 
 Nevertheless, at the quality of N-body data, the model predictions  
 based on the ECM and the SCM quantitatively agree with N-body results  
 in cases with spectral index $n<0$. 
 For the index $n\ge0$, choice of the Lagrangian local dynamics becomes 
 crucial for an accurate prediction and a more delicate modeling is 
 required, however, we find that the model prediction based on the ECM 
 provides a better approximation to the N-body results of cumulants and PDFs. 
\end{abstract}
%%%%%%%%%%%%%%%%%%%%%%%%%%%%%%%%%%%%%%%%%%%%%%%%%%%%%%%%%%%%%%%%%%%%%%
%
%
%
%
%
%
%
%
%
%%%%%%%%%%%%%%%%%%%%%%%%%%%%%%%%%%%%%%%%%%%%%%%%%%%%%%%%%%%%%%%%%%%%%%
%%%%%%%%%%%%%%%%%%%%%%%%%%%%%%%%%%%%%%%%%%%%%%%%%%%%%%%%%%%%%%%%%%%%%%
\section{Introduction}\label{sec:intro}
%%%%%%%%%%%%%%%%%%%%%%%%%%%%%%%%%%%%%%%%%%%%%%%%%%%%%%%%%%%%%%%%%%%%%%
%%%%%%%%%%%%%%%%%%%%%%%%%%%%%%%%%%%%%%%%%%%%%%%%%%%%%%%%%%%%%%%%%%%%%%
The probability distribution function (PDF) of the cosmological density
fluctuation is a fundamental statistical quantity characterizing the
large-scale structure of the universe. In a standard picture of cosmic
structure formation based on the cold dark matter scenario, the
gravitational evolution of the dark matter distribution plays an
essential role for the hierarchical nature of observed luminous
distributions. Usually, the evolution of dark matter distribution is
believed to be developed from a small initial fluctuation with Gaussian
random distribution. While the PDF of density fluctuation retains
Gaussian shape in a linear regime, the deviation from Gaussian
distribution becomes significant in the non-linear regime of
gravitational evolution.

A number of studies in quantifying the non-Gaussian properties of
density field have been developed theoretically and
observationally. From the numerical and the observational study, a
systematic analysis using the cosmological N-body simulation or the
observed galaxy distribution yield various phenomenological prescription
for the density PDF in the non-linear regime
\citep[e.g.,][]{SH1984,H1985,GY1993,UY1996}. Among them, the lognormal
distribution has been long known to fit to the simulations quite
accurately \citep[e.g.,][]{CJ1991,CMS1993,BK1995,TW2000}. Recently,
\citet{KTS2001} critically examined this issue using the high resolution
N-body simulation with Gaussian initial conditions, and found that the
accuracy of the lognormal model remains valid, irrespective of the
nature of initial spectra. The weak dependence of the initial spectra
was later investigated using the phenomenological models with dark halo
approach \citep{THK2003}.

On the other hand, from the analytical study, a perturbative
construction of the PDFs has been exploited by \citet{B1992,B1994a}
employing a field-theoretical approach and the predictions including the
smoothing effect excellently match the N-body simulations in the weakly
non-linear regime. Beyond the perturbative prediction, however, no exact
treatment is present and the non-perturbative approximation or the
phenomenological approach taking account of the empirical simulation
results are necessary. \citet{FG1998} and \citet{SG2001} proposed to use
a spherical collapse model as a non-perturbative approximation to
predict the higher-order moments and PDFs. In their treatment, one
assumes that the Lagrangian dynamics of the local density field is
simply described by spherical collapse model. Although this
approximation clearly misses the non-locality of the gravity in the
sense that the evolution of the local density field can be determined by
the one-to-one local mapping, the advantage of this treatment is that
one can easily calculate the higher-order correction of the moments and
PDFs. Further, it turns out that the spherical collapse approximation
exactly recovers the leading order results of perturbation theory.

Recently, we generalize the idea of spherical collapse approximation to
the local approximation in which the evolution of local density field is
characterized by the Lagrangian local dynamics with finite degrees of
freedom \citep{OKT2003}. As a demonstration, the PDFs were computed using
the ellipsoidal collapse model. In the ellipsoidal collapse model, the
local density at a position is expressed as the multivariate function of
initial parameters, i.e., the principal axes of the ellipsoid given at
the same position. Thus, the relation between the initial and evolved
density field cannot be described by the one-to-one local mapping. As a
consequence, the local approximation with ellipsoidal collapse model
successfully explains the stochastic nature seen in the simulation,
i.e., the joint probability between the initial and the evolved density
fields, as has been reported by \citet{KTS2001}. In addition, the
leading-order results from the ellipsoidal collapse model correctly reproduce
those obtained from the exact perturbation theory.

In the present paper, we extend the previous study to the quantitative
comparison between the local approximation and the N-body
simulations. Evaluating the PDF of local density fields taking account
of the smoothing effect, we consider the validity and the limitation of
the local approximation with spherical and ellipsoidal collapse
models. The PDFs from the spherical collapse model were previously
compared with the N-body simulations in the case with cold dark matter
(CDM) power spectrum \citep{SG2001}. In this paper, taking account of
the smoothing effect, the N-body results with the scale-free initial
spectra as well as the CDM spectrum are compared.

This paper is organized as follows. In section \ref{sec:LCM}, we start
to review the local approximation of one-point statistics developed by
\citet{OKT2003} and briefly show how to compute the PDF and the moments
from the Lagrangian local collapse model. As representative models of
Lagrangian local dynamics, the spherical and the ellipsoidal collapse
model are considered. Then, we consider the smoothing effect and discuss
how to incorporate it into the model predictions. Based on this, 
the perturbative calculation of cumulants up to the two-loop order 
is presented and the qualitative behaviors of the model prediction are
discussed in section \ref{sec:perturbation_ECM}. In section 
\ref{sec:results}, the validity and the usefulness of the local approximation 
for one-point statistics is investigated by comparing the 
PDFs and cumulants from the local collapse models 
with those obtained from the N-body simulations. 
Finally, section
\ref{sec:conclusion} is devoted to discussion and conclusions.
%
%
%
%
%
%
%
%
%
%%%%%%%%%%%%%%%%%%%%%%%%%%%%%%%%%%%%%%%%%%%%%%%%%%%%%%%%%%%%%%%%%%%%%%
%%%%%%%%%%%%%%%%%%%%%%%%%%%%%%%%%%%%%%%%%%%%%%%%%%%%%%%%%%%%%%%%%%%%%%
\section{One-point statistics from the local collapse model}
\label{sec:LCM}
%%%%%%%%%%%%%%%%%%%%%%%%%%%%%%%%%%%%%%%%%%%%%%%%%%%%%%%%%%%%%%%%%%%%%%
%%%%%%%%%%%%%%%%%%%%%%%%%%%%%%%%%%%%%%%%%%%%%%%%%%%%%%%%%%%%%%%%%%%%%%
%
%
%
%
%
%
%
%
%
In many analytical works on the gravitational evolution of density
distributions, the cold dark matter distribution is often treated
as the pressureless and non-relativistic fluid. This treatment 
is not exact, but in a statistical sense, it would provide a better approximation if the scale of our 
interest is large enough, where no shell-crossing appears in the smoothed 
density fields. Denoting the mass density and 
velocity field of fluid by $\rho$ and $\bm{v}$, the evolution equations
for the fluid in a homogeneous and isotropic background universe are
expressed as follows:
%%%%%%%%%%%%%%%%%%%%%%%%%%%%%%%%%%%%%%%%%%%%%%%%%%%%%%%%%%%%%%%%%%%%%%
\begin{equation}
 \frac{\partial\delta}{\partial t}+\frac{1}{a}\nabla\cdot[(1+\delta)\bm{v}]=0,
  \label{eq:continuity}
\end{equation}
%%%%%%%%%%%%%%%%%%%%%%%%%%%%%%%%%%%%%%%%%%%%%%%%%%%%%%%%%%%%%%%%%%%%%%
\begin{equation}
 \frac{\partial\bm{v}}{\partial t}+H\bm{v}+\frac{1}{a}(\bm{v}\cdot\nabla)\bm{v}=-\frac{1}{a}\nabla\phi,
\end{equation}
%%%%%%%%%%%%%%%%%%%%%%%%%%%%%%%%%%%%%%%%%%%%%%%%%%%%%%%%%%%%%%%%%%%%%%
where $\delta$ is the density fluctuation,
$\delta\equiv(\rho-\rho_{\rm m})/\rho_{\rm m}$. The quantity $a$ is the
scale factor of the universe and $H$ is the Hubble parameter given by
$H\equiv\dot{a}/a$. Gravitational potential $\phi$ is determined by the
Poisson equation:
%%%%%%%%%%%%%%%%%%%%%%%%%%%%%%%%%%%%%%%%%%%%%%%%%%%%%%%%%%%%%%%%%%%%%%
\begin{equation}
 \nabla^2\phi=4\pi G\rho_{\rm m}a^2\delta.\label{eq:poisson}
\end{equation}
%%%%%%%%%%%%%%%%%%%%%%%%%%%%%%%%%%%%%%%%%%%%%%%%%%%%%%%%%%%%%%%%%%%%%%

Below, owing to the local approximation, we consider the one-point PDF
of the density fluctuation $\delta$ and the higher-order moments taking
account of the smoothing effect.
%
%
%
%
%
%
%%%%%%%%%%%%%%%%%%%%%%%%%%%%%%%%%%%%%%%%%%%%%%%%%%%%%%%%%%%%%%%%%%%%%%
\subsection{Local approximation}
\label{subsec:local}
%%%%%%%%%%%%%%%%%%%%%%%%%%%%%%%%%%%%%%%%%%%%%%%%%%%%%%%%%%%%%%%%%%%%%%
%
%
%
%
%
As mentioned in section \ref{sec:intro}, the analytical treatment of the
one-point PDF $P(\delta)$ governed by the fluid equations
(\ref{eq:continuity})-(\ref{eq:poisson}) is generally intractable due to
the non-linearity and the non-locality of the gravity. Beyond the
perturbative prediction, a non-perturbative treatment should be
exploited. The local approximation is one of the way to treat the
one-point PDFs analytically by reducing the Lagrangian dynamics of the
fluid motion given by (\ref{eq:continuity})-(\ref{eq:poisson}) to the
local dynamics with finite degrees of freedom. In this treatment, the
time evolution of the local density field $\delta$ at a given position is
determined by the local dynamics with the initial condition given at the
same position in Lagrangian coordinate. Thus, the solution of local
density field can be obtained by solving a couple of ordinary
differential equation and expressed as a function of initial parameters
$\bm{\lambda}=(\lambda_1,\lambda_2,\cdots,\lambda_n)$ given at 
a Lagrangian position and time, i.e.,
$\delta=f(\bm{\lambda},t)$. In this paper, we consider the spherical and
the ellipsoidal collapse models as representative example of the
Lagrangian local dynamics (see section \ref{subsubsec:SCM} and
\ref{subsubsec:ECM}). In this case, the initial parameters of Lagrangian
local dynamics correspond to the linearly extrapolated density
fluctuation $\delta_l$ for the spherical collapse model and the
principal axes of initial homogeneous ellipsoid,
$(\lambda_1,\lambda_2,\lambda_3)$ for the ellipsoidal collapse model.

Once provided the functional form of the local density
$f(\bm{\lambda},t)$, the one-point PDF of the local density field
$P(\delta;t)$ (in Eulerian space) can be analytically obtained. 
With a slight modification
of the definition of density fluctuation $\delta$ so as to satisfy the
normalization condition and the zero-mean of $\delta$ \citep{OKT2003},
one has
%%%%%%%%%%%%%%%%%%%%%%%%%%%%%%%%%%%%%%%%%%%%%%%%%%%%%%%%%%%%%%%%%%%%%%
\begin{equation}
 P(\delta,t)=\frac{1}{1+\delta}\int \prod_{i=1}^n d\lambda_i 
  \probI(\bm{\lambda})\delta_D[\delta-g(\bm{\lambda},t)],
  \label{eq:EulPDFdelta}
\end{equation}
%%%%%%%%%%%%%%%%%%%%%%%%%%%%%%%%%%%%%%%%%%%%%%%%%%%%%%%%%%%%%%%%%%%%%%
where the function $g$ is given by
%%%%%%%%%%%%%%%%%%%%%%%%%%%%%%%%%%%%%%%%%%%%%%%%%%%%%%%%%%%%%%%%%%%%%%
\begin{equation}
 \delta=g(\bm{\lambda},t)\equiv N_E[1+f(\bm{\lambda},t)]-1~~;\,\,\,
  N_E(t)\equiv\int \prod_{i=1}^n d\lambda_i\frac{\probI(\bm{\lambda})}{1+f(\bm{\lambda},t)}.
  \label{eq:function_g}
\end{equation}
%%%%%%%%%%%%%%%%%%%%%%%%%%%%%%%%%%%%%%%%%%%%%%%%%%%%%%%%%%%%%%%%%%%%%%
In equations (\ref{eq:EulPDFdelta}) and (\ref{eq:function_g}), the
function $\probI(\bm{\lambda})$ is the probability distribution of the
initial parameters, which characterizes the randomness of the mass
distribution. From (\ref{eq:EulPDFdelta}), one also calculates the
moments of the density fields:
%%%%%%%%%%%%%%%%%%%%%%%%%%%%%%%%%%%%%%%%%%%%%%%%%%%%%%%%%%%%%%%%%%%%%%
\begin{equation}
 \left\langle\delta^N\right\rangle \equiv \int \delta^N P(\delta,t)d\delta
  = \int\prod_{i=1}^n d\lambda_i \frac{g^N}{1+g} \probI(\bm{\lambda}).
  \label{eq:NonSmoothMoment}
\end{equation}
%%%%%%%%%%%%%%%%%%%%%%%%%%%%%%%%%%%%%%%%%%%%%%%%%%%%%%%%%%%%%%%%%%%%%%
The expressions (\ref{eq:EulPDFdelta})-(\ref{eq:NonSmoothMoment}) are
the heart of the analytical treatment in local approximation, which are
rigorously derived by considering the evolution equations for the
one-point PDFs \citep{OKT2003}.
%
%
%
%
%
%%%%%%%%%%%%%%%%%%%%%%%%%%%%%%%%%%%%%%%%%%%%%%%%%%%%%%%%%%%%%%%%%%%%%%
\subsubsection{Spherical collapse model}
\label{subsubsec:SCM}
%%%%%%%%%%%%%%%%%%%%%%%%%%%%%%%%%%%%%%%%%%%%%%%%%%%%%%%%%%%%%%%%%%%%%%
%
%
%
%
%
As simple Lagrangian local dynamics to calculate the function
$f(\bm{\lambda},t)$, let us first consider the spherical collapse
model (SCM). In the SCM, the evolution of local density at a given
position in Lagrangian space is determined by the mass $M$ inside a
sphere of radius $R$ collapsing via self gravity:
%%%%%%%%%%%%%%%%%%%%%%%%%%%%%%%%%%%%%%%%%%%%%%%%%%%%%%%%%%%%%%%%%%%%%%
\begin{equation}
 \frac{d^2R}{dt^2}=-\frac{GM}{R^2}+\frac{\Lambda}{3}R\quad;\quad
 M=\frac{4\pi}{3}\bar{\rho}R^3=\mbox{const},\label{eq:SCevol}
\end{equation}
%%%%%%%%%%%%%%%%%%%%%%%%%%%%%%%%%%%%%%%%%%%%%%%%%%%%%%%%%%%%%%%%%%%%%%
where $\Lambda$ is cosmological constant. The above equation is
re-expressed in terms of the local density defined by
$\delta=(aR_0/R)^3-1$:
%%%%%%%%%%%%%%%%%%%%%%%%%%%%%%%%%%%%%%%%%%%%%%%%%%%%%%%%%%%%%%%%%%%%%%
\begin{equation}
 \frac{d^2\delta}{dt^2}+2H\frac{d\delta}{dt}-
 \frac{4}{3}\frac{1}{1+\delta}\left(\frac{d\delta}{dt}
\right)^2=\frac{3}{2}H^2\Omega_{\rm m}(1+\delta)\delta,
\end{equation}
%%%%%%%%%%%%%%%%%%%%%%%%%%%%%%%%%%%%%%%%%%%%%%%%%%%%%%%%%%%%%%%%%%%%%%
where $\Omega_{\rm m}$ is the density parameter given by
$\Omega_{\rm m}\equiv8\pi G\rho_{\rm m}/(3H^2)$. Note that the SCM can
be regarded as the monopole approximation of the fluid equations
neglecting the shear and vorticity \citep[e.g.][]{FG1998}, since one obtains 
the following
equation from (\ref{eq:continuity})-(\ref{eq:poisson}) with a help of
the Lagrangian time derivative,
$d/dt\equiv\partial/\partial t+\bm{v}/a\cdot\nabla$:
%%%%%%%%%%%%%%%%%%%%%%%%%%%%%%%%%%%%%%%%%%%%%%%%%%%%%%%%%%%%%%%%%%%%%%
\begin{equation}
 \frac{d^2\delta}{dt^2}+2H\frac{d\delta}{dt}-\frac{4}{3}\frac{1}{1+\delta}\left(\frac{d\delta}{dt}
\right)^2=H^2(1+\delta)\left(\frac{3}{2}\Omega_{\rm m}\delta+\sigma^{ij}\sigma_{ij}-\omega^{ij}\omega_{ij}
\right);
\label{eq:fluid_eq_Lag}
\end{equation}
%%%%%%%%%%%%%%%%%%%%%%%%%%%%%%%%%%%%%%%%%%%%%%%%%%%%%%%%%%%%%%%%%%%%%%
%%%%%%%%%%%%%%%%%%%%%%%%%%%%%%%%%%%%%%%%%%%%%%%%%%%%%%%%%%%%%%%%%%%%%%
\begin{eqnarray}
 \sigma_{ij}&=&\frac{1}{2aH}\left(\frac{\partial v_i}{\partial x_j}
  +\frac{\partial v_j}{\partial x_i}
  -\frac{2}{3}\nabla\cdot\bm{v}\delta_{ij}\right),\\
 \omega_{ij}&=&\frac{1}{2aH}\left(\frac{\partial v_i}{\partial x_j}
  -\frac{\partial v_j}{\partial x_i}\right),
\end{eqnarray}
%%%%%%%%%%%%%%%%%%%%%%%%%%%%%%%%%%%%%%%%%%%%%%%%%%%%%%%%%%%%%%%%%%%%%%
where $\sigma_{ij}$ and $\omega_{ij}$ respectively denote shear and
vorticity tensor.

The exact solution of the equation (\ref{eq:SCevol}) is obtained in the
case of the Einstein-de Sitter universe ($\Omega_{\rm m}=1,\Lambda=0$)
and is expressed as the function of linearly extrapolated density
fluctuation $\delta_l$ in parametric form:
%%%%%%%%%%%%%%%%%%%%%%%%%%%%%%%%%%%%%%%%%%%%%%%%%%%%%%%%%%%%%%%%%%%%%%
\begin{equation}
 \delta=\frac{9}{2}\frac{(\eta-\sin\eta)^2}{(1-\cos\eta)^3}-1,\quad\delta_l=\frac{3}{5}\left[\frac{3}{4}(\eta-\sin\eta)\right]^{2/3}\label{eq:SCsolution+}
\end{equation}
%%%%%%%%%%%%%%%%%%%%%%%%%%%%%%%%%%%%%%%%%%%%%%%%%%%%%%%%%%%%%%%%%%%%%%
for $\delta_l>0$, and
%%%%%%%%%%%%%%%%%%%%%%%%%%%%%%%%%%%%%%%%%%%%%%%%%%%%%%%%%%%%%%%%%%%%%%
\begin{equation}
 \delta=\frac{9}{2}\frac{(\sinh\eta-\eta)^2}{(\cosh\eta-1)^3}-1,\quad\delta_l=-\frac{3}{5}\left[\frac{3}{4}(\sinh\eta-\eta)\right]^{2/3}\label{eq:SCsolution-}
\end{equation}
%%%%%%%%%%%%%%%%%%%%%%%%%%%%%%%%%%%%%%%%%%%%%%%%%%%%%%%%%%%%%%%%%%%%%%
for $\delta_l<0$. The relation between $\delta$ and $\delta_l$ in the
above equation is fairly accurate even in the non Einstein-de Sitter
universe \citep[e.g.][]{NS1995,FG1998b}. We will extensively use
(\ref{eq:SCsolution+}) and
(\ref{eq:SCsolution-}) for later analysis. Note that in computing the
density PDF, the linearly extrapolated density $\delta_l$ should be
regarded as initial parameter and is treated as a random
variable. Assuming the Gaussian initial condition, we have
%%%%%%%%%%%%%%%%%%%%%%%%%%%%%%%%%%%%%%%%%%%%%%%%%%%%%%%%%%%%%%%%%%%%%%
\begin{equation}
 \probI(\delta_l)=\frac{1}{\sqrt{2\pi}\sigma_l}e^{-(\delta_l/\sigma_l)^2/2},
\end{equation}
%%%%%%%%%%%%%%%%%%%%%%%%%%%%%%%%%%%%%%%%%%%%%%%%%%%%%%%%%%%%%%%%%%%%%%
where the variable $\sigma_l$ is the rms fluctuation of linear density
field, $\sigma_l=\langle\delta_l^2\rangle^{1/2}$.
%
%
%
%
%
%
%
%%%%%%%%%%%%%%%%%%%%%%%%%%%%%%%%%%%%%%%%%%%%%%%%%%%%%%%%%%%%%%%%%%%%%%
\subsubsection{Ellipsoidal collapse model}\label{subsubsec:ECM}
%%%%%%%%%%%%%%%%%%%%%%%%%%%%%%%%%%%%%%%%%%%%%%%%%%%%%%%%%%%%%%%%%%%%%%
%
%
%
%
%
The ellipsoidal collapse model (ECM) is an extension of the SCM taking
account of the non-sphericity. In this model, the dynamics of the local
density is described by the self-gravitating uniform density ellipsoid
characterized by the half length of the principal axes
$\alpha_i$ ($i=1,2,3$). 

The local density of ECM is given by
%%%%%%%%%%%%%%%%%%%%%%%%%%%%%%%%%%%%%%%%%%%%%%%%%%%%%%%%%%%%%%%%%%%%%%
\begin{equation}
 \delta=\frac{a^3}{\alpha_1\alpha_2\alpha_3}-1.
  \label{eq:delta_ellipsoid}
\end{equation}
%%%%%%%%%%%%%%%%%%%%%%%%%%%%%%%%%%%%%%%%%%%%%%%%%%%%%%%%%%%%%%%%%%%%%%
According to \citet{BM1996}, the evolution equations of half length of
axes become
%%%%%%%%%%%%%%%%%%%%%%%%%%%%%%%%%%%%%%%%%%%%%%%%%%%%%%%%%%%%%%%%%%%%%%
\begin{equation}
 \frac{d^2}{dt^2}\alpha_i= \frac{\Lambda}{3}\alpha_i
  -4\pi G\,\rho_{\rm m}\,\alpha_i
  \left(\frac{1+\delta}{3}+\frac{b_i}{2}\delta +\lambda_{{\rm ext},i}\right),
  \label{eq:alpha_evol}
\end{equation}
%%%%%%%%%%%%%%%%%%%%%%%%%%%%%%%%%%%%%%%%%%%%%%%%%%%%%%%%%%%%%%%%%%%%%%
\begin{equation}
 b_i=\alpha_1\alpha_2\alpha_3\int_0^\infty
  \frac{d\tau}{(\alpha_i^2+\tau)\prod_j(\alpha_j^2+\tau)^{1/2}}-\frac{2}{3}.
  \label{eq:carlson}
\end{equation}
%%%%%%%%%%%%%%%%%%%%%%%%%%%%%%%%%%%%%%%%%%%%%%%%%%%%%%%%%%%%%%%%%%%%%%
The quantity
$\lambda_{{\rm ext},i}$ mimics the effect of external tidal shear, which
was introduced for the consistency with Zel'dovich
approximation:
%%%%%%%%%%%%%%%%%%%%%%%%%%%%%%%%%%%%%%%%%%%%%%%%%%%%%%%%%%%%%%%%%%%%%%
\begin{equation}
 \lambda_{{\rm ext},i}=\cases{\displaystyle 
  \lambda_i-\frac{\lambda_1+\lambda_2+\lambda_3}{3} &; 
  linear external tide , \cr
  \displaystyle \frac{5}{4}\,b_i &; non-linear external tide ,\cr}
  \label{eq:external_tide}
\end{equation}
%%%%%%%%%%%%%%%%%%%%%%%%%%%%%%%%%%%%%%%%%%%%%%%%%%%%%%%%%%%%%%%%%%%%%%
where $\lambda_i$ denotes the initial perturbation of principal axis and
evolves as $\lambda_i(t)=D(t)\lambda_i(t_0)$ with the variable $D$ being
the linear growth rate. Note that the inclusion of the external tidal term 
is necessary to reproduce the Zel'dovich approximation when we linearize 
the evolution equations.\footnote{If this term is dropped, a consistent 
calculation with initial distribution (\ref{eq:DoroPDF})  is also impossible.}
In this paper, both cases of the external tidal
term are considered in order to reveal the model dependence of the
prediction, but in comparing with N-body simulation, the ECM results with 
linear external tide is presented for brevity. 
In presence of the external tide, the initial condition is 
specified by the Zel'dovich approximation. Identifying the variables
$\lambda_i$ with the initial parameters of $\alpha_i$, we have
%%%%%%%%%%%%%%%%%%%%%%%%%%%%%%%%%%%%%%%%%%%%%%%%%%%%%%%%%%%%%%%%%%%%%%
\begin{eqnarray}
 \alpha_i(t_0)&=&a(t_0)[1-\lambda_i(t_0)]\label{eq:init_alpha}\\
 \frac{d}{dt}\alpha_i(t_0)&=&\dot{a}(t_0)[1-\lambda_i(t_0)]-a(t_0)\dot{\lambda}_i(t_0).\label{eq:init_alpha_dt}
\end{eqnarray}
%%%%%%%%%%%%%%%%%%%%%%%%%%%%%%%%%%%%%%%%%%%%%%%%%%%%%%%%%%%%%%%%%%%%%%
Note that the initial parameters $\lambda_i$ are related to the
linearly extrapolated density fluctuation $\delta_l$ 
by $\delta_l=\lambda_1+\lambda_2+\lambda_3$. Hence, the parameters
$\lambda_i$ should be treated as the random variables. For the Gaussian
initial condition, the distribution function of $\lambda_i$ is
analytically expressed as follows \citep{D1970}:
%%%%%%%%%%%%%%%%%%%%%%%%%%%%%%%%%%%%%%%%%%%%%%%%%%%%%%%%%%%%%%%%%%%%%%
\begin{equation}
 \probI(\lambda_i)=\frac{675\sqrt{5}}{8\pi\sigma_l^6}
  \exp\left(-3\frac{I_1^2}{\sigma_l^2}+15\frac{I_2}{2\sigma_l^2}\right)
  (\lambda_1-\lambda_2)(\lambda_2-\lambda_3)(\lambda_1-\lambda_3),
  \label{eq:DoroPDF}
\end{equation}
%%%%%%%%%%%%%%%%%%%%%%%%%%%%%%%%%%%%%%%%%%%%%%%%%%%%%%%%%%%%%%%%%%%%%%
where we define $I_1=\lambda_1+\lambda_2+\lambda_3$,
$I_2=\lambda_1\lambda_2+\lambda_2\lambda_3+\lambda_3\lambda_1$.

Note that in contrast to the SCM, the ECM can be
regarded as an approximation of the fluid equations taking account of
the effect of tidal shear but neglecting the vorticity \citep{OKT2003}. 
The equation (\ref{eq:alpha_evol}) with (\ref{eq:carlson}) is
re-expressed in terms of the local density $\delta$ 
(c.f., eq.[\ref{eq:fluid_eq_Lag}]):
%%%%%%%%%%%%%%%%%%%%%%%%%%%%%%%%%%%%%%%%%%%%%%%%%%%%%%%%%%%%%%%%%%%%%%
\begin{equation}
\frac{d^2\delta}{dt^2}+2H\frac{d\delta}{dt}
 -\frac{4}{3}\frac{1}{1+\delta}\left(\frac{d\delta}{dt}\right)^2
 =H^2(1+\delta)\left(\frac{3}{2}\Omega_{\rm m}\delta
		+\sigma^{ij}\sigma_{ij}\right);\label{eq:eom_delta_ellipsoid}
\end{equation}
%%%%%%%%%%%%%%%%%%%%%%%%%%%%%%%%%%%%%%%%%%%%%%%%%%%%%%%%%%%%%%%%%%%%%%
\begin{equation}
 \sigma_{ij}=\frac{1}{3H}\left(3\frac{\dot{\alpha_i}}{\alpha_i}
   -\frac{\dot{\alpha _1}}{\alpha_1}-\frac{\dot{\alpha_2}}{\alpha_2}
   -\frac{\dot{\alpha_3}}{\alpha_3}\right)\delta_{ij}.
\end{equation}
%%%%%%%%%%%%%%%%%%%%%%%%%%%%%%%%%%%%%%%%%%%%%%%%%%%%%%%%%%%%%%%%%%%%%%
%
%
%
%
%
%%%%%%%%%%%%%%%%%%%%%%%%%%%%%%%%%%%%%%%%%%%%%%%%%%%%%%%%%%%%%%%%%%%%%%
\subsection{Smoothing effect}
%%%%%%%%%%%%%%%%%%%%%%%%%%%%%%%%%%%%%%%%%%%%%%%%%%%%%%%%%%%%%%%%%%%%%%
%
%
%
%
%
When we evaluate the statistical quantities from the N-body simulations,
the smoothing procedure is often employed to remedy the discreteness of
the particle data. In this sense, the smoothing effect is crucial
and should be incorporated into the theoretical prediction. In this
paper, we adopt the top-hat smoothing and the smoothed density PDFs are
computed from both the local approximation and N-body simulation:
%%%%%%%%%%%%%%%%%%%%%%%%%%%%%%%%%%%%%%%%%%%%%%%%%%%%%%%%%%%%%%%%%%%%%%
\begin{eqnarray}
 \hat{\delta}(\bm{x};R)&=&\int d^3y\,\delta(\bm{y})
  W_{\rm TH}(|\bm{y}-\bm{x}|;R)\,\,;\\
 W_{\rm TH}(r;R)&=&\left\{ \begin{array}{ll}
		            3/4\pi R^3 & r<R \\
			    0 & \mbox{otherwise}
			   \end{array}, \right.
\end{eqnarray}
%%%%%%%%%%%%%%%%%%%%%%%%%%%%%%%%%%%%%%%%%%%%%%%%%%%%%%%%%%%%%%%%%%%%%%
where $W_{\rm TH}(r;R)$ is the top-hat smoothing kernel of the radius
$R$.

A systematic method to compute the analytic PDFs taking account of the
smoothing effect was first considered by \citet{B1994a} based on the
perturbation theory. Later, his method was extended to the
non-perturbative calculation of the PDF using SCM \citep{FG1998}. We
briefly review the method by \citet{FG1998}.

First notice the fact that in the case of the top-hat smoothing, the
leading-order results of cumulants for local density field is not
affected by the smoothing in Lagrangian space
\citep{B1994a}. Extrapolating this result to the non-perturbative
approximation with SCM, one can approximate the evolved local density
with top-hat smoothing by
%%%%%%%%%%%%%%%%%%%%%%%%%%%%%%%%%%%%%%%%%%%%%%%%%%%%%%%%%%%%%%%%%%%%%%
\begin{equation}
 \hat{\delta} \approx f(\delta_{l,\rm L}),
  \label{eq:smoothLag}
\end{equation}
%%%%%%%%%%%%%%%%%%%%%%%%%%%%%%%%%%%%%%%%%%%%%%%%%%%%%%%%%%%%%%%%%%%%%%
where $\hat{\delta}$ is the smoothed density and $\delta_{l,\rm L}$ is
the linear density fluctuation in Lagrangian space.
To relate the quantity $\delta_{l,\rm L}$ with that in Eulerian space,
$\delta_{l,\rm E}$, we recall the fact that the radius $R$ defined in
the Eulerian space roughly corresponds to the radius
$R(1+\hat{\delta})^{1/3}$ in Lagrangian space. Thus, we have
%%%%%%%%%%%%%%%%%%%%%%%%%%%%%%%%%%%%%%%%%%%%%%%%%%%%%%%%%%%%%%%%%%%%%%
 \begin{equation}
  \frac{\delta_{l,\rm E}}{\sigma_l(R)}=
   \frac{\delta_{l,\rm L}}{\sigma_l[R(1+\hat{\delta})^{1/3}]}.
   \label{eq:relEulLag}
 \end{equation}
%%%%%%%%%%%%%%%%%%%%%%%%%%%%%%%%%%%%%%%%%%%%%%%%%%%%%%%%%%%%%%%%%%%%%%
Substituting (\ref{eq:relEulLag}) into (\ref{eq:smoothLag}) and
identifying the linear fluctuation $\delta_{l,\rm E}$ with the smoothed
linear fluctuation $\hat{\delta}_l$, the relation between $\hat{\delta}$
and $\hat{\delta}_l$ becomes
%%%%%%%%%%%%%%%%%%%%%%%%%%%%%%%%%%%%%%%%%%%%%%%%%%%%%%%%%%%%%%%%%%%%%%
 \begin{equation}
  \hat{\delta}=\hat{f}(\hat{\delta}_l)\approx
   f\left\{\frac{\sigma_l[R(1+\hat{\delta})^{1/3}]}{\sigma_l(R)}
     \hat{\delta}_l\right\}.
   \label{eq:sphsmooth_prime}
 \end{equation}
%%%%%%%%%%%%%%%%%%%%%%%%%%%%%%%%%%%%%%%%%%%%%%%%%%%%%%%%%%%%%%%%%%%%%%
The above relation can be further simplified by the running index
$\gamma$ defined by $\gamma \equiv d\log \sigma_l^2/d\log R$:
%%%%%%%%%%%%%%%%%%%%%%%%%%%%%%%%%%%%%%%%%%%%%%%%%%%%%%%%%%%%%%%%%%%%%%
 \begin{equation}
 \hat{f}(\hat{\delta}_l)=f\left\{\left[1+\hat{f}(\hat{\delta}_l)\right]^{\gamma/6}\hat{\delta}_l\right\}\quad;\quad\gamma = \frac{d\log \sigma_l^2}{d\log R} = -(n+3),\label{eq:sphsmooth}
 \end{equation}
%%%%%%%%%%%%%%%%%%%%%%%%%%%%%%%%%%%%%%%%%%%%%%%%%%%%%%%%%%%%%%%%%%%%%%
where $n$ is the index of the initial power spectrum,
$P(k)\propto k^n$. Notice that the above expression is still valid in
the non power-law cases of the initial power spectrum. The remarkable
feature in the relation (\ref{eq:sphsmooth_prime}) or
(\ref{eq:sphsmooth}) is that the leading order result of the cumulants
obtained from the perturbation theory is exactly recovered by the local
approximation with SCM.

Owing to this noticeable fact, we extend to use the result
(\ref{eq:sphsmooth}) to the local approximation with the ECM. In this
case, the smoothed density field is related to the top-hat filtered
principal axis $\hat{\lambda}_i$ given by:
%%%%%%%%%%%%%%%%%%%%%%%%%%%%%%%%%%%%%%%%%%%%%%%%%%%%%%%%%%%%%%%%%%%%%%
\begin{eqnarray}
 \hat{f}(\hat{\bm{\lambda}},t)&=&f\left\{\left[1+\hat{f}(\hat{\bm{\lambda}},t)\right]^{\gamma/6}\hat{\bm{\lambda}},t\right\},\label{eq:delta_smooth}
\end{eqnarray}
%%%%%%%%%%%%%%%%%%%%%%%%%%%%%%%%%%%%%%%%%%%%%%%%%%%%%%%%%%%%%%%%%%%%%%
Adopting this form, the PDF (\ref{eq:EulPDFdelta}) becomes
%%%%%%%%%%%%%%%%%%%%%%%%%%%%%%%%%%%%%%%%%%%%%%%%%%%%%%%%%%%%%%%%%%%%%%
\begin{equation}
  P(\hat{\delta};R)=\frac{1}{1+\hat{\delta}}\int\prod_{i=1}^3 d\hat{\lambda}_i \probI(\hat{\bm{\lambda}})\delta_D\left[\hat{\delta}-\hat{g}(\hat{\bm{\lambda}},t)\right],\label{eq:smoothedDc}
\end{equation}
%%%%%%%%%%%%%%%%%%%%%%%%%%%%%%%%%%%%%%%%%%%%%%%%%%%%%%%%%%%%%%%%%%%%%%
with $\hat{g}(\hat{\bm{\lambda}},t)$ being
%%%%%%%%%%%%%%%%%%%%%%%%%%%%%%%%%%%%%%%%%%%%%%%%%%%%%%%%%%%%%%%%%%%%%%
\begin{equation}
 \hat{g}(\hat{\bm{\lambda}},t)=\hat{N}_E\left[1+\hat{f}(\hat{\bm{\lambda}}),t\right]-1\quad;\quad
\hat{N}_E(t)=\int\prod_i d\hat{\lambda}_i \frac{\probI(\hat{\bm{\lambda}})}{1+\hat{f}(\hat{\bm{\lambda}},t)}.
\label{eq: g_and_N}
\end{equation}
%%%%%%%%%%%%%%%%%%%%%%%%%%%%%%%%%%%%%%%%%%%%%%%%%%%%%%%%%%%%%%%%%%%%%%
The above equations apparently seem difficult to evaluate because of the
implicit relation of the functions $\hat{f}$ and
$\hat{\bm{\lambda}}$. However, this apparent difficulty can be eliminated by
introducing the following variables:
%%%%%%%%%%%%%%%%%%%%%%%%%%%%%%%%%%%%%%%%%%%%%%%%%%%%%%%%%%%%%%%%%%%%%%
\begin{equation}
 \lambda_i'=\left[1+\hat{f}(\hat{\bm{\lambda}},t)\right]^{\gamma/6}
  \hat{\lambda}_i.
\end{equation}
%%%%%%%%%%%%%%%%%%%%%%%%%%%%%%%%%%%%%%%%%%%%%%%%%%%%%%%%%%%%%%%%%%%%%%
Then the equation (\ref{eq:smoothedDc}) becomes
%%%%%%%%%%%%%%%%%%%%%%%%%%%%%%%%%%%%%%%%%%%%%%%%%%%%%%%%%%%%%%%%%%%%%%
\begin{equation}
  P(\hat{\delta};R)=\frac{1}{1+\hat{\delta}}\int\prod_{i=1}^3 d\lambda_i' 
  P_I\left\{\left[1+f(\bm{\lambda}',t)\right]^{-\gamma/6}\bm{\lambda}'\right\}
  \delta_D\left[\hat{\delta}-\hat{g}(\bm{\lambda}',t)\right]
  \left|\frac{\partial\hat{\lambda}_j}{\partial\lambda_k'}\right|
  \label{eq:smoothPDFd},
\end{equation}
%%%%%%%%%%%%%%%%%%%%%%%%%%%%%%%%%%%%%%%%%%%%%%%%%%%%%%%%%%%%%%%%%%%%%%
where the quantities $\hat{g}$ and $\hat{N}_E$ can be recast as
%%%%%%%%%%%%%%%%%%%%%%%%%%%%%%%%%%%%%%%%%%%%%%%%%%%%%%%%%%%%%%%%%%%%%%
\begin{equation}
 \hat{g}(\bm{\lambda}',t) = \hat{N}_E\left[1+f(\bm{\lambda}',t)\right]-1
  ~~;\quad \hat{N}_E(t)=\int\prod_i d\lambda_i' 
  \frac{P_I\left\{\left[1+f(\bm{\lambda}',t)\right]^{-\gamma/6}\bm{\lambda}'\right\}}{1+f(\bm{\lambda}',t)}
  \left|\frac{\partial\hat{\lambda}_j}{\partial\lambda_k'}\right|.
\end{equation}
%%%%%%%%%%%%%%%%%%%%%%%%%%%%%%%%%%%%%%%%%%%%%%%%%%%%%%%%%%%%%%%%%%%%%%
Also, the Jacobian $|\partial\hat{\lambda}_j/\partial\lambda_k'|$ is
calculated using the relation
$\hat{\lambda}_i=[1+f(\bm{\lambda},t)]^{-\gamma/6}\lambda_i'$ as:
%%%%%%%%%%%%%%%%%%%%%%%%%%%%%%%%%%%%%%%%%%%%%%%%%%%%%%%%%%%%%%%%%%%%%%
\begin{equation}
 \left|\frac{\partial\hat{\lambda}_j}{\partial\lambda_k'}\right|=\left[1+f(\bm{\lambda}',t)\right]^{-\gamma/2}\left(1-\frac{\gamma}{6}\frac{1}{1+f}\sum_{i=1}^{3}\lambda_i'\frac{\partial f}{\partial \lambda_i'}\right).
  \label{eq:Jacobian}
\end{equation}
%%%%%%%%%%%%%%%%%%%%%%%%%%%%%%%%%%%%%%%%%%%%%%%%%%%%%%%%%%%%%%%%%%%%%%
Note that the term
$-1/(1+f)\sum_{i=1}^{3}\lambda_i'\,(\partial f/\partial\lambda_i')$ is
related to the velocity divergence and is expressed as
$\theta=\sum_i\dot{\alpha}_i/(H\alpha_i)-3$ in the case of the
Einstein-de Sitter universe. Thus, provided the solution of the
equations for ECM (eqs.[\ref{eq:delta_ellipsoid}]-[\ref{eq:carlson}]),
the smoothed density PDF can be numerically evaluated by substituting
the solution $f(\bm{\lambda},t)$ into the above equations. Similarly,
the higher-order moments of the local density becomes
%%%%%%%%%%%%%%%%%%%%%%%%%%%%%%%%%%%%%%%%%%%%%%%%%%%%%%%%%%%%%%%%%%%%%%
\begin{equation}
 \langle \hat{\delta}^N\rangle = \int \hat{\delta}^N P(\hat{\delta};R)d\delta 
  = \int \prod_{i=1}^3d\lambda'_i\frac{\hat{g}^N}{1+\hat{g}}\probI\left[(1+f)^{-\gamma/6}\bm{\lambda}'\right]\left| \frac{\partial\hat{\lambda}_j}{\partial\lambda'_k}\right|.
  \label{eq:MomentFromPdf}
\end{equation}
%%%%%%%%%%%%%%%%%%%%%%%%%%%%%%%%%%%%%%%%%%%%%%%%%%%%%%%%%%%%%%%%%%%%%%

In what follows, we simply denote the smoothed density field
$\hat{\delta}$ by $\delta$.
%
%
%
%
%
%
%
%
%
%
%
%
%
%%%%%%%%%%%%%%%%%%%%%%%%%%%%%%%%%%%%%%%%%%%%%%%%%%%%%%%%%%%%%%%%%%%%%%
%%%%%%%%%%%%%%%%%%%%%%%%%%%%%%%%%%%%%%%%%%%%%%%%%%%%%%%%%%%%%%%%%%%%%%
\section{Perturbation theory in Ellipsoidal Collapse Model}
\label{sec:perturbation_ECM}
%%%%%%%%%%%%%%%%%%%%%%%%%%%%%%%%%%%%%%%%%%%%%%%%%%%%%%%%%%%%%%%%%%%%%%
%%%%%%%%%%%%%%%%%%%%%%%%%%%%%%%%%%%%%%%%%%%%%%%%%%%%%%%%%%%%%%%%%%%%%%
%
%
%
%
%
%
%
%
%
Before proceeding to compare the theoretical models with N-body simulations, 
it is useful to examine the perturbative analysis of the local collapse models. 
In this section, based on the ellipsoidal collapse model, 
we present the perturbative calculation of the cumulants 
of density field. The differences between the model predictions as well as the qualitative 
behaviors are also discussed in section \ref{subsubsec:comparison}.

Here and in what follows, we treat the evolution of local density in an 
Einstein-de
Sitter universe. In this case, the linear growth rate $D$ is simply
proportional to the scale factor $a$. Let us denote the half-length of
the principal axis $\alpha_i$ by
%%%%%%%%%%%%%%%%%%%%%%%%%%%%%%%%%%%%%%%%%%%%%%%%%%%%%%%%%%%%%%%%%%%%%%
\begin{equation}
\alpha_i(t) = a(t)\{1-\zeta_i(a(t))\}.
\label{eq:def_of_zeta}
\end{equation}
%%%%%%%%%%%%%%%%%%%%%%%%%%%%%%%%%%%%%%%%%%%%%%%%%%%%%%%%%%%%%%%%%%%%%%
Then, regarding the scale factor $a$ as time variable, 
the equation (\ref{eq:alpha_evol}) is transformed to 
%%%%%%%%%%%%%%%%%%%%%%%%%%%%%%%%%%%%%%%%%%%%%%%%%%%%%%%%%%%%%%%%%%%%%%
\begin{equation}
 a^2\frac{d^2\zeta_i}{da^2}+\frac{3}{2}a\frac{d\zeta_i}{da}=
\frac{3}{2}(1-\zeta_i)\left(\frac{1}{3}\delta+
\frac{1}{2}b_i\delta+\lambda_{{\rm ext},i}\right).
\label{eq:ellip_zeta}
\end{equation}
%%%%%%%%%%%%%%%%%%%%%%%%%%%%%%%%%%%%%%%%%%%%%%%%%%%%%%%%%%%%%%%%%%%%%%
In terms of the variable $\zeta_i$, the quantities $b_i$ and $\delta$ are 
expressed as 
%%%%%%%%%%%%%%%%%%%%%%%%%%%%%%%%%%%%%%%%%%%%%%%%%%%%%%%%%%%%%%%%%%%%%%
\begin{eqnarray}
 b_i&=&(1-\zeta_1)(1-\zeta_2)(1-\zeta_3)\int_0^\infty
  \frac{d\tau}{\{(1-\zeta_i)^2+\tau\}\prod_j\{(1-\zeta_j)^2+\tau\}^{1/2}}
  -\frac{2}{3} 
\label{eq:b_i}
\\
  &\equiv&\frac{4}{15}\{3\zeta_i-(\zeta_1+\zeta_2+\zeta_3)\}+\tilde{b}_i,\\
\delta&\equiv&\zeta_1+\zeta_2+\zeta_3+\Delta.
\end{eqnarray}
%%%%%%%%%%%%%%%%%%%%%%%%%%%%%%%%%%%%%%%%%%%%%%%%%%%%%%%%%%%%%%%%%%%%%%
Note that both of the quantities $\tilde{b}_i$ and $\Delta$ are of the
order of ${\cal O}(\zeta^2)$.

Below, we separately give the perturbation results in the ellipsoidal
collapse model with non-linear external tide and with linear external
tide. Note that the leading-order results for the cumulants in each
model exactly coincide with each other and reproduce those of the exact
perturbation theory \citep[e.g.,][]{B1994a,B1994b}.
%
%
%
%
%
%
%
%
%
%%%%%%%%%%%%%%%%%%%%%%%%%%%%%%%%%%%%%%%%%%%%%%%%%%%%%%%%%%%%%%%%%%%%%%
%%%%%%%%%%%%%%%%%%%%%%%%%%%%%%%%%%%%%%%%%%%%%%%%%%%%%%%%%%%%%%%%%%%%%%
\subsection{Ellipsoidal collapse model with non-linear external tide}
\label{subsubsec:nonlinear-tide}
%%%%%%%%%%%%%%%%%%%%%%%%%%%%%%%%%%%%%%%%%%%%%%%%%%%%%%%%%%%%%%%%%%%%%%
%%%%%%%%%%%%%%%%%%%%%%%%%%%%%%%%%%%%%%%%%%%%%%%%%%%%%%%%%%%%%%%%%%%%%%
%
%
%
%
%
%
%
%
%
In the case of the model with non-linear external tide, the quantity
$\lambda_{{\rm ext},i}$ is given by $(5/4)b_i$ and the right hand side
of the equation (\ref{eq:ellip_zeta}) becomes
$3\zeta_i/2+{\cal O}(\zeta^2)$. Then We have  
%%%%%%%%%%%%%%%%%%%%%%%%%%%%%%%%%%%%%%%%%%%%%%%%%%%%%%%%%%%%%%%%%%%%%%
\begin{equation}
 a^2\frac{d^2\zeta_i}{da^2}+\frac{3}{2}a\frac{d\zeta_i}{da}-\frac{3}{2}\zeta_i=\frac{1}{2}(\Delta-\delta\zeta_i)+\frac{3}{4}(1-\zeta_i)b_i\delta+\frac{15}{8}\left(\tilde{b}_i-\zeta_ib_i\right)={\cal O}(\zeta^2). 
\end{equation}
%%%%%%%%%%%%%%%%%%%%%%%%%%%%%%%%%%%%%%%%%%%%%%%%%%%%%%%%%%%%%%%%%%%%%%
This equation can be perturbatively solved by substituting the series
expansion $\zeta_i=\sum_{j=0}\xi_i^{(j)}a^j$ into the above
equation. Under the initial condition (\ref{eq:init_alpha}), one
formally obtains the expression of the coefficient $\xi_i^{(j)}$ as
%%%%%%%%%%%%%%%%%%%%%%%%%%%%%%%%%%%%%%%%%%%%%%%%%%%%%%%%%%%%%%%%%%%%%%
\begin{eqnarray}
 \xi_i^{(1)}&=&\lambda_i(t_0)\\
 \xi_i^{(j)}&=&\!\!\!\frac{1}{(2j+3)(j-1)j!}\,\frac{d^j}{da^j}\!\left.\left[
\Delta-\delta\zeta_i+\frac{3}{2}(1-\zeta_i)b_i\delta+\frac{15}{4}\left(\tilde{b}_i-\zeta_ib_i\right)
\right]\right|_{a=0};\, j>1, 
\label{eq:solution_zeta_i}
\end{eqnarray}
%%%%%%%%%%%%%%%%%%%%%%%%%%%%%%%%%%%%%%%%%%%%%%%%%%%%%%%%%%%%%%%%%%%%%%
where we have only considered the growing mode of the solutions. Note
that the coefficient $\xi_i^{(j)}$ is the variable of the order of
${\cal O}(\lambda^j)$.

Based on the result (\ref{eq:solution_zeta_i}), the perturbative
expansion for the evolved density, $\delta=\sum_i\,\delta^{(i)}$, 
can be constructed from 
the relations (\ref{eq:delta_ellipsoid}), (\ref{eq:def_of_zeta}) and
(\ref{eq:b_i}). The results up to the seventh order become
%%%%%%%%%%%%%%%%%%%%%%%%%%%%%%%%%%%%%%%%%%%%%%%%%%%%%%%%%%%%%%%%%%%%%%
\begin{eqnarray}
 \delta^{(1)}&=&\delta_l,
\\
 \delta^{(2)}&=&\frac{17}{21}\delta_l^2+\frac{4}{21}J_1,
\\
 \delta^{(3)}&=&\frac{341}{567}\delta_l^3+\frac{338}{945}\delta_lJ_1
  +\frac{92}{3969}J_2,
\\
 \delta^{(4)}&=&\frac{55805}{130977}\delta_l^4
  +\frac{485288}{1091475}\delta_l^2J_1+\frac{234088}{4584195}\delta_lJ_2
  +\frac{429728}{10696455}J_1^2,
\\
 \delta^{(5)}&=&\frac{213662}{729729}\delta_l^5
  +\frac{292398464}{638512875}\delta_l^3J_1
  +\frac{64182728}{893918025}\delta_l^2J_2
\nonumber \\
 & &\ +\frac{6541246}{59594535}\delta_lJ_1^2
  +\frac{828974992}{96364363095}J_1J_2,
\\
 \delta^{(6)}&=& \frac{21129781}{107270163}\delta_l^6
  +\frac{15739030628}{37246584375}\delta_l^4J_1
  +\frac{38380501904}{469306963125}\delta_l^3J_2
\nonumber\\
 & &\ +\frac{334168450808}{1825082634375}\delta_l^2J_1^2
  +\frac{250313183728}{9368757523125}\delta_lJ_1J_2
\nonumber\\
 & &\ +\frac{63778345006048}{7673012411439375}J_1^3
  +\frac{2272657750768}{4603807446863625}J_2^2,
\\
 \delta^{(7)}&=&\frac{83411812}{639441621}\delta_l^7
  +\frac{42267062029204}{116564878828125}\delta_l^5J_1
  +\frac{63224677073056}{769328200265625}\delta_l^4J_2
\nonumber\\
 & &\ +\frac{3019799334120902}{12565693937671875}\delta_l^3J_1^2
  +\frac{9581060236980928}{193511686640146875}\delta_l^2J_1J_2
\nonumber\\
 & &\ +\frac{13515215809239748}{451527268827009375}\delta_lJ_1^3
  +\frac{276922264619192}{162549816777723375}\delta_lJ_2^2
\nonumber\\
 & &\ +\frac{9155185965341512}{3521912696850673125}J_1^2J_2, 
\end{eqnarray}
%%%%%%%%%%%%%%%%%%%%%%%%%%%%%%%%%%%%%%%%%%%%%%%%%%%%%%%%%%%%%%%%%%%%%%
where $\delta_l$ denotes the linear fluctuation given by
$\delta_l=\lambda_1+\lambda_2+\lambda_3$. Here, we introduced the
quantities $J_1\equiv x^2+xy+y^2$ and $J_2\equiv (x-y)(2x+y)(x+2y)$ with
the variables $x$ and $y$ being $x=\lambda_1-\lambda_2$ and
$y=\lambda_2-\lambda_3$, respectively.

Once provided the perturbative solution for the non-smoothed density field,
cumulants for the smoothed density is calculated as follows.
From the perturbative inversion of the relation (\ref{eq:delta_smooth}),  
the smoothed density $\hat{\delta}$ is obtained and the normalization factor 
$\hat{N}_E$ is first calculated by substituting this into the definition
(\ref{eq: g_and_N}).
Using the probability distribution of
initial parameter (\ref{eq:DoroPDF}), the resultant expression becomes 
%%%%%%%%%%%%%%%%%%%%%%%%%%%%%%%%%%%%%%%%%%%%%%%%%%%%%%%%%%%%%%%%%%%%%%
\begin{eqnarray}
 N_E&=&1-\frac{1}{6}\gamma\sigma_l^2
  +\left(\frac{10844}{848925}-\frac{79}{4410}\gamma-\frac{31}{378}\gamma^2
    -\frac{1}{27}\gamma^3\right)\sigma_l^4\nonumber\\
 &&+\left(\frac{3891599696}{511023137625}-\frac{1248901}{278107830}\gamma
     -\frac{47093}{415800}\gamma^2\right.\nonumber\\
 &&~~\left.-\frac{62341}{317520}\gamma^3-\frac{19}{168}\gamma^4
      -\frac{1}{48}\gamma^5\right)\sigma_l^6 
\end{eqnarray}
%%%%%%%%%%%%%%%%%%%%%%%%%%%%%%%%%%%%%%%%%%%%%%%%%%%%%%%%%%%%%%%%%%%%%%
up to the order ${\cal O}(\sigma_l^6)$. Then, the moments for the
smoothed density 
$\langle\hat{\delta}^N\rangle$ is evaluated from the relation 
$\langle\hat{\delta}^N\rangle=\int \prod_i\, d\lambda_i\,\, \hat{g}^N/(1+\hat{g})\,
\probI(\bm{\lambda})$ (c.f. eq.[\ref{eq:NonSmoothMoment}]). Finally, the
perturbative
correction for the variance $\sigma^2=\langle\hat{\delta}^2\rangle$, the
skewness $S_3=\langle\hat{\delta}^3\rangle/\sigma^4$ and the kurtosis
$S_4=(\langle\hat{\delta}^4\rangle-3\sigma^4)/\sigma^6$ are obtained and can
be summarized as series expansion of $\sigma_l^2$ as follows: 
%%%%%%%%%%%%%%%%%%%%%%%%%%%%%%%%%%%%%%%%%%%%%%%%%%%%%%%%%%%%%%%%%%%%%%%
\begin{eqnarray}
 \sigma^2&=&\sigma_l^2 + s_{2,4}\,\sigma_l^4 + s_{2,6}\,\sigma_l^6 + 
s_{2,8}\,\sigma_l^8 + \cdots, 
\label{eq:perturbation_sigma}\\
 S_3&=& S_{3,0} + S_{3,2}\,\sigma_l^2 + S_{3,4}\,\sigma_l^4+\cdots,
\label{eq:perturbation_S3}\\
 S_4&=& S_{4,0} + S_{4,2}\,\sigma_l^2 + S_{4,4}\,\sigma_l^4+\cdots.
\label{eq:perturbation_S4}
\end{eqnarray}
%%%%%%%%%%%%%%%%%%%%%%%%%%%%%%%%%%%%%%%%%%%%%%%%%%%%%%%%%%%%%%%%%%%%%%%
The resultant expressions for the coefficients $s_{2,i}$, $S_{3,i}$ and 
$S_{4,i}$ become 
%%%%%%%%%%%%%%%%%%%%%%%%%%%%%%%%%%%%%%%%%%%%%%%%%%%%%%%%%%%%%%%%%%%%%%
\begin{eqnarray}
 s_{2,4}&=&\frac{439}{245}+\frac{167}{126}\gamma+\frac{11}{36}\gamma^2,
\\
 s_{2,6}&=&\frac{3143785639}{695269575}+\frac{15856223}{2037420}\gamma
  +\frac{55273}{10584}\gamma^2+\frac{1835}{1134}\gamma^3
  +\frac{127}{648}\gamma^4,
\\
 s_{2,8}&=&\frac{7932609222047169799}{537532462889296875}
  +\frac{2321384486861437}{54752479031250}\gamma
  +\frac{1062497682871}{20858087250}\gamma^2\nonumber\\
 &&~~+\frac{16268385923}{495093060}\gamma^3+\frac{61875775}{5143824}\gamma^4
  +\frac{13831}{5832}\gamma^5+\frac{6877}{34992}\gamma^6
\end{eqnarray}
%%%%%%%%%%%%%%%%%%%%%%%%%%%%%%%%%%%%%%%%%%%%%%%%%%%%%%%%%%%%%%%%%%%%%%
for the variance,  
%%%%%%%%%%%%%%%%%%%%%%%%%%%%%%%%%%%%%%%%%%%%%%%%%%%%%%%%%%%%%%%%%%%%%%
\begin{eqnarray}
 S_{3,0}&=&\frac{34}{7}+\gamma,
\\
 S_{3,2}&=&\frac{1041064}{101871}
  +\frac{21946}{2205}\gamma+\frac{415}{126}\gamma^2+\frac{10}{27}\gamma^3,
\\
 S_{3,4}&=&\frac{161751288183332}{3041804390625}
  +\frac{363349617641}{3476347875}\gamma
  +\frac{415283963}{5093550}\gamma^2\nonumber\\
 &&~~+\frac{11299781}{357210}\gamma^3+\frac{2975}{486}\gamma^4
  +\frac{1841}{3888}\gamma^5
\end{eqnarray}
%%%%%%%%%%%%%%%%%%%%%%%%%%%%%%%%%%%%%%%%%%%%%%%%%%%%%%%%%%%%%%%%%%%%%%
for the skewness and 
%%%%%%%%%%%%%%%%%%%%%%%%%%%%%%%%%%%%%%%%%%%%%%%%%%%%%%%%%%%%%%%%%%%%%%
\begin{eqnarray}
 S_{4,0}&=&\frac{60712}{1323}+\frac{62}{3}\gamma+\frac{7}{3}\gamma^2,
\\
 S_{4,2}&=&\frac{941370178286}{3476347875}+\frac{1518808496}{4584195}\gamma
  +\frac{18161033}{119070}\gamma^2+\frac{3935}{126}\gamma^3
  +\frac{1549}{648}\gamma^4,
\\
 S_{4,4}&=&\frac{30144942925392628918}{13782883663828125}
  +\frac{129392230965050887}{27376239515625}\gamma
  +\frac{397096017904379}{93861392625}\gamma^2\nonumber\\
 &&~~+\frac{30126971437}{15002820}\gamma^3
  +\frac{1142621801}{2143260}\gamma^4+\frac{6127195}{81648}\gamma^5
  +\frac{102005}{23328}\gamma^6
\end{eqnarray}
%%%%%%%%%%%%%%%%%%%%%%%%%%%%%%%%%%%%%%%%%%%%%%%%%%%%%%%%%%%%%%%%%%%%%%
for the kurtosis. 
%
%
%
%
%
%
%
%
%
%%%%%%%%%%%%%%%%%%%%%%%%%%%%%%%%%%%%%%%%%%%%%%%%%%%%%%%%%%%%%%%%%%%%%%
%%%%%%%%%%%%%%%%%%%%%%%%%%%%%%%%%%%%%%%%%%%%%%%%%%%%%%%%%%%%%%%%%%%%%%
\subsection{Ellipsoidal collapse model with linear external tide }
\label{subsubsec:linear-tide}
%%%%%%%%%%%%%%%%%%%%%%%%%%%%%%%%%%%%%%%%%%%%%%%%%%%%%%%%%%%%%%%%%%%%%%
%%%%%%%%%%%%%%%%%%%%%%%%%%%%%%%%%%%%%%%%%%%%%%%%%%%%%%%%%%%%%%%%%%%%%%
%
%
%
%
%
%
%
%
%
For the perturbative solution in the model with the linear external tide, 
the calculation is slightly reduced if we introduce the quantities 
$A=\zeta_1+\zeta_2+\zeta_3$, $B=\zeta_1-\zeta_2$ and $C=\zeta_2-\zeta_3$. 
Recalling the fact that the external tidal term becomes 
$\lambda_{{\rm ext},i}=\lambda_i-(\lambda_1+\lambda_2+\lambda_3)/3$, 
the evolution equation (\ref{eq:ellip_zeta}) is rewritten with 
%%%%%%%%%%%%%%%%%%%%%%%%%%%%%%%%%%%%%%%%%%%%%%%%%%%%%%%%%%%%%%%%%%%%%%
\begin{eqnarray}
 a^2\frac{d^2A}{da^2}+\frac{3}{2}a\frac{dA}{da}-\frac{3}{2}A&=&\frac{3}{2}\Delta-\frac{1}{2}\delta A-\frac{3}{4}\sum_{i=1}^{3}\zeta_i(\delta b_i+2\lambda_{{\rm ext},i}),
\label{eq:eom_A}\\
 a^2\frac{d^2B}{da^2}+\frac{3}{2}a\frac{dB}{da}-\frac{3}{2}(\lambda_1-\lambda_2)&=&\frac{3}{4}\delta\{(1-\zeta_1)b_1-(1-\zeta_2)b_2\}-\frac{1}{2}\delta B
\nonumber\\
 &&-\frac{3}{2}(\zeta_1\lambda_{{\rm ext},1}-\zeta_2\lambda_{{\rm ext},2}),
\label{eq:eom_B}\\
a^2\frac{d^2C}{da^2}+\frac{3}{2}a\frac{dC}{da}-\frac{3}{2}(\lambda_2-\lambda_3)&=&\frac{3}{4}\delta\{(1-\zeta_2)b_2-(1-\zeta_3)b_3\}-\frac{1}{2}\delta C
\nonumber\\ 
 &&-\frac{3}{2}(\zeta_2\lambda_{{\rm ext},2}-\zeta_3\lambda_{{\rm ext},3}).
\label{eq:eom_C}
\end{eqnarray}
%%%%%%%%%%%%%%%%%%%%%%%%%%%%%%%%%%%%%%%%%%%%%%%%%%%%%%%%%%%%%%%%%%%%%%
Note also that the right hand sides of the equation 
(\ref{eq:eom_A})-(\ref{eq:eom_C})  are of the order of ${\cal O}(\lambda^2)$. 

Similar to the procedure in section \ref{subsubsec:nonlinear-tide}, 
the perturbative expansion for the density $\delta$ is constructed 
from the perturbative solutions of $A$, $B$ and $C$. 
After a tedious but a straightforward calculation, 
the perturbative solutions up to the seventh order become 
%%%%%%%%%%%%%%%%%%%%%%%%%%%%%%%%%%%%%%%%%%%%%%%%%%%%%%%%%%%%%%%%%%%%%%
\begin{eqnarray}
 \delta^{(1)}&=&\delta_l,
\\
 \delta^{(2)}&=& \frac{17}{21}\delta_l^2+\frac{4}{21}J_1,
\\
 \delta^{(3)}&=&\frac{341}{567}\delta_l^3+\frac{1538}{4725}\delta_lJ_1
  +\frac{4}{405}J_2,
\\
 \delta^{(4)}&=&\frac{55805}{130977}\delta_l^4
  +\frac{952144}{2480625}\delta_l^2J_1+\frac{345088}{16372125}\delta_lJ_2
  +\frac{12368}{363825}J_1^2,
\\
\delta^{(5)}&=&\frac{213662}{729729}\delta_l^5
 +\frac{237342074}{621928125}\delta_l^3J_1
 +\frac{93363344}{3192564375}\delta_l^2J_2
\nonumber \\
 & &\ +\frac{52865818}{638512875}\delta_lJ_1^2+\frac{135052}{34827975}J_1J_2,
\\
 \delta^{(6)}&=&\frac{21129781}{107270163}\delta_l^6
  +\frac{73816004896012}{215099024765625}\delta_l^4J_1
  +\frac{29134959410408}{879950555859375}\delta_l^3J_2
\nonumber\\
 & &\ +\frac{45534497984}{355535578125}\delta_l^2J_1^2
  +\frac{1416570594232}{129059414859375}\delta_lJ_1J_2
\nonumber\\
 & &\ +\frac{797014912}{132368630625}J_1^3
  +\frac{162352}{1578740625}J_2^2,
\\
 \delta^{(7)}&=&\frac{83411812}{639441621}\delta_l^7
  +\frac{24700151148166244}{85566392051765625}\delta_l^5J_1
\nonumber\\
 & &\ +\frac{49159400006961656}{1480956785511328125}\delta_l^4J_2
  +\frac{18569055254261594}{116681443706953125}\delta_l^3J_1^2
\nonumber\\
 & &\ +\frac{13101172588796}{684531136414125}\delta_l^2J_1J_2
  +\frac{9323215177292}{488950811724375}\delta_lJ_1^3
\nonumber\\
 & &\ +\frac{1452480606736}{4400557305519375}\delta_lJ_2^2
  +\frac{11003633175272}{10267967046211875}J_1^2J_2.
\end{eqnarray}
%%%%%%%%%%%%%%%%%%%%%%%%%%%%%%%%%%%%%%%%%%%%%%%%%%%%%%%%%%%%%%%%%%%%%%
The normalization factor of the PDF is 
%%%%%%%%%%%%%%%%%%%%%%%%%%%%%%%%%%%%%%%%%%%%%%%%%%%%%%%%%%%%%%%%%%%%%%
\begin{eqnarray}
 N_E&=&1-\frac{1}{6}\gamma\sigma_l^2
  +\left(\frac{69668}{3898125}
    +\frac{701}{198450}\gamma-\frac{31}{378}\gamma^2
    -\frac{1}{27}\gamma^4\right)\sigma_l^4 \nonumber\\
 &&+\left(\frac{5033872069084}{645297074296875}
     +\frac{3873942169}{223479506250}\gamma
     -\frac{38616157}{509355000}\gamma^2\right.\nonumber\\
 &&~~~\left.-\frac{286169}{1587600}\gamma^3
       -\frac{19}{168}\gamma^4-\frac{1}{48}\gamma^5\right)\sigma_l^6.
\end{eqnarray}
%%%%%%%%%%%%%%%%%%%%%%%%%%%%%%%%%%%%%%%%%%%%%%%%%%%%%%%%%%%%%%%%%%%%%%
Then, the coefficients of the perturbative correction for cumulants become
%%%%%%%%%%%%%%%%%%%%%%%%%%%%%%%%%%%%%%%%%%%%%%%%%%%%%%%%%%%%%%%%%%%%%%
\begin{eqnarray}
 s_{2,4}&=&\frac{57137}{33075}+\frac{167}{126}\gamma+\frac{11}{36}\gamma^2,
\\
 s_{2,6}&=&\frac{469828713881}{111739753125}
  +\frac{17130160379}{2292097500}\gamma+\frac{488945}{95256}\gamma^2
  +\frac{1835}{1134}\gamma^3+\frac{127}{648}\gamma^4,
\\
 s_{2,8}&=&\frac{996244294855051546571}{74870593045294921875}
  +\frac{152613969392185373}{3871782445781250}\gamma
  +\frac{683964582869801}{14079208893750}\gamma^2
\nonumber\\
 &&~~~+\frac{1980638022487}{61886632500}\gamma^3
  +\frac{306289019}{25719120}\gamma^4+\frac{13831}{5832}\gamma^5
  +\frac{6877}{34992}\gamma^6
\end{eqnarray}
%%%%%%%%%%%%%%%%%%%%%%%%%%%%%%%%%%%%%%%%%%%%%%%%%%%%%%%%%%%%%%%%%%%%%%
for the variance,  
%%%%%%%%%%%%%%%%%%%%%%%%%%%%%%%%%%%%%%%%%%%%%%%%%%%%%%%%%%%%%%%%%%%%%%
\begin{eqnarray}
 S_{3,0}&=&\frac{34}{7}+\gamma,
\\
 S_{3,2}&=&\frac{646404856}{63669375}+\frac{327062}{33075}\gamma
  +\frac{415}{126}\gamma^2+\frac{10}{27}\gamma^3,
\\
 S_{3,4}&=&\frac{77881923244216108}{1505693173359375}
     +\frac{80186055186641}{782178271875}\gamma
     +\frac{2052918391}{25467750}\gamma^2
\nonumber\\
 &&~~~+\frac{56239289}{1786050}\gamma^3
  +\frac{2975}{486}\gamma^4+\frac{1841}{3888}\gamma^5
\end{eqnarray}
%%%%%%%%%%%%%%%%%%%%%%%%%%%%%%%%%%%%%%%%%%%%%%%%%%%%%%%%%%%%%%%%%%%%%%
for the skewness and 
%%%%%%%%%%%%%%%%%%%%%%%%%%%%%%%%%%%%%%%%%%%%%%%%%%%%%%%%%%%%%%%%%%%%%%
\begin{eqnarray}
 S_{4,0}&=&\frac{60712}{1323}+\frac{62}{3}\gamma+\frac{7}{3}\gamma^2,
\\
 S_{4,2}&=&\frac{210688932175742}{782178271875}
  +\frac{188859083824}{573024375}\gamma+\frac{90600877}{595350}\gamma^2
  +\frac{3935}{126}\gamma^3+\frac{1549}{648}\gamma^4,
\\
 S_{4,4}&=&\frac{102133149992759420855618}{47644922847005859375}
  +\frac{7007053550215029257}{1505693173359375}\gamma
\nonumber\\
 &&~~+\frac{29469590927547677}{7039604446875}\gamma^2
  +\frac{13726819977457}{6876292500}\gamma^3
\nonumber\\
 &&~~+\frac{5700735749}{10716300}\gamma^4+\frac{6127195}{81648}\gamma^5
  +\frac{102005}{23328}\gamma^6
\end{eqnarray}
%%%%%%%%%%%%%%%%%%%%%%%%%%%%%%%%%%%%%%%%%%%%%%%%%%%%%%%%%%%%%%%%%%%%%%
for the kurtosis. 
%
%
%
%
%
%
%
%
%%%%%%%%%%%%%%%%%%%%%%%%%%%%%%%%%%%%%%%%%%%%%%%%%%%%%%%%%%%%%%%%%%%%%%
%%%%%%%%%%%%%%%%%%%%%%%%%%%%%%%%%%%%%%%%%%%%%%%%%%%%%%%%%%%%%%%%%%%%%%
\subsection{Differences between model predictions}
\label{subsubsec:comparison}
\begin{figure}
 \plottwo{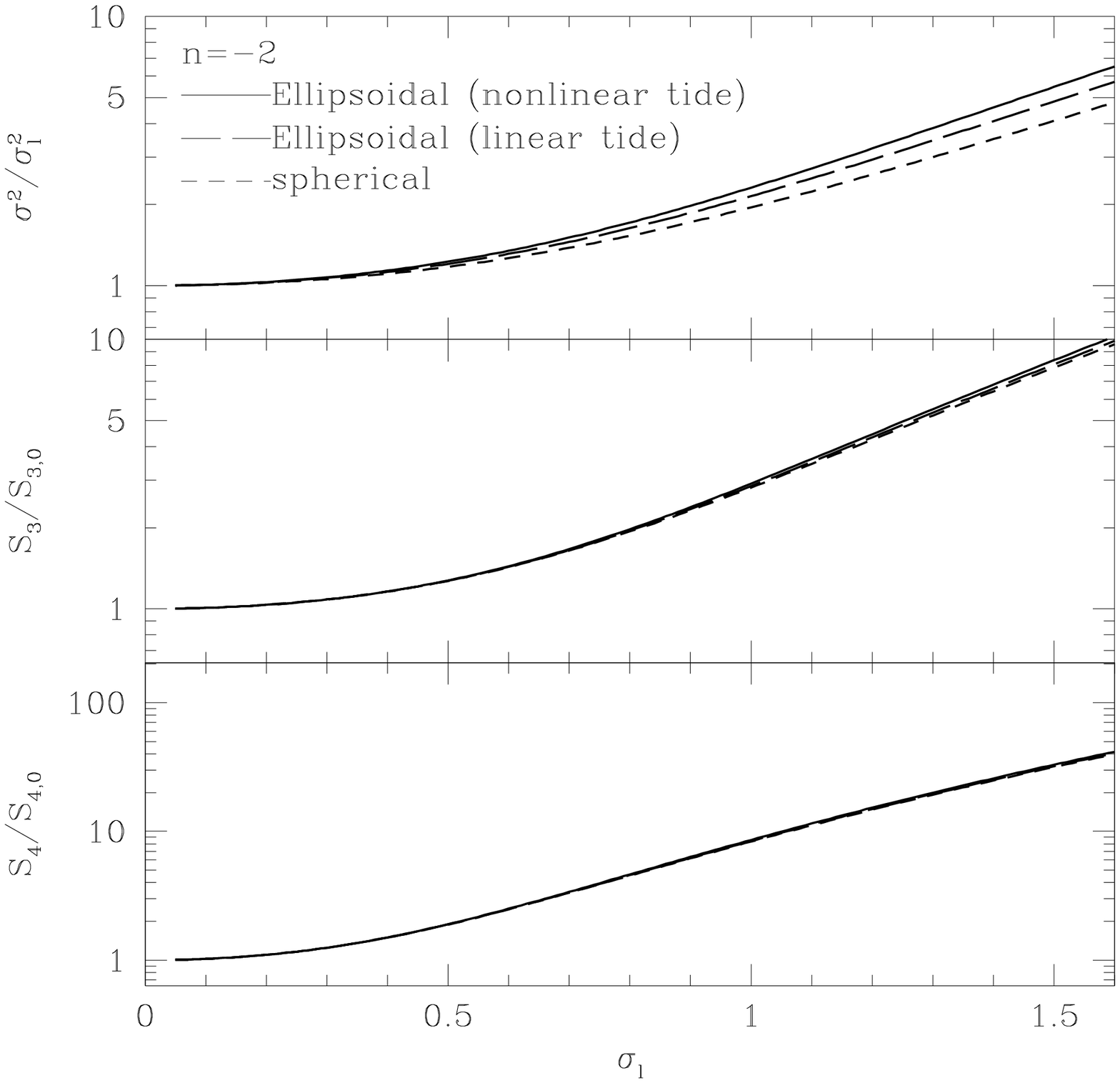}{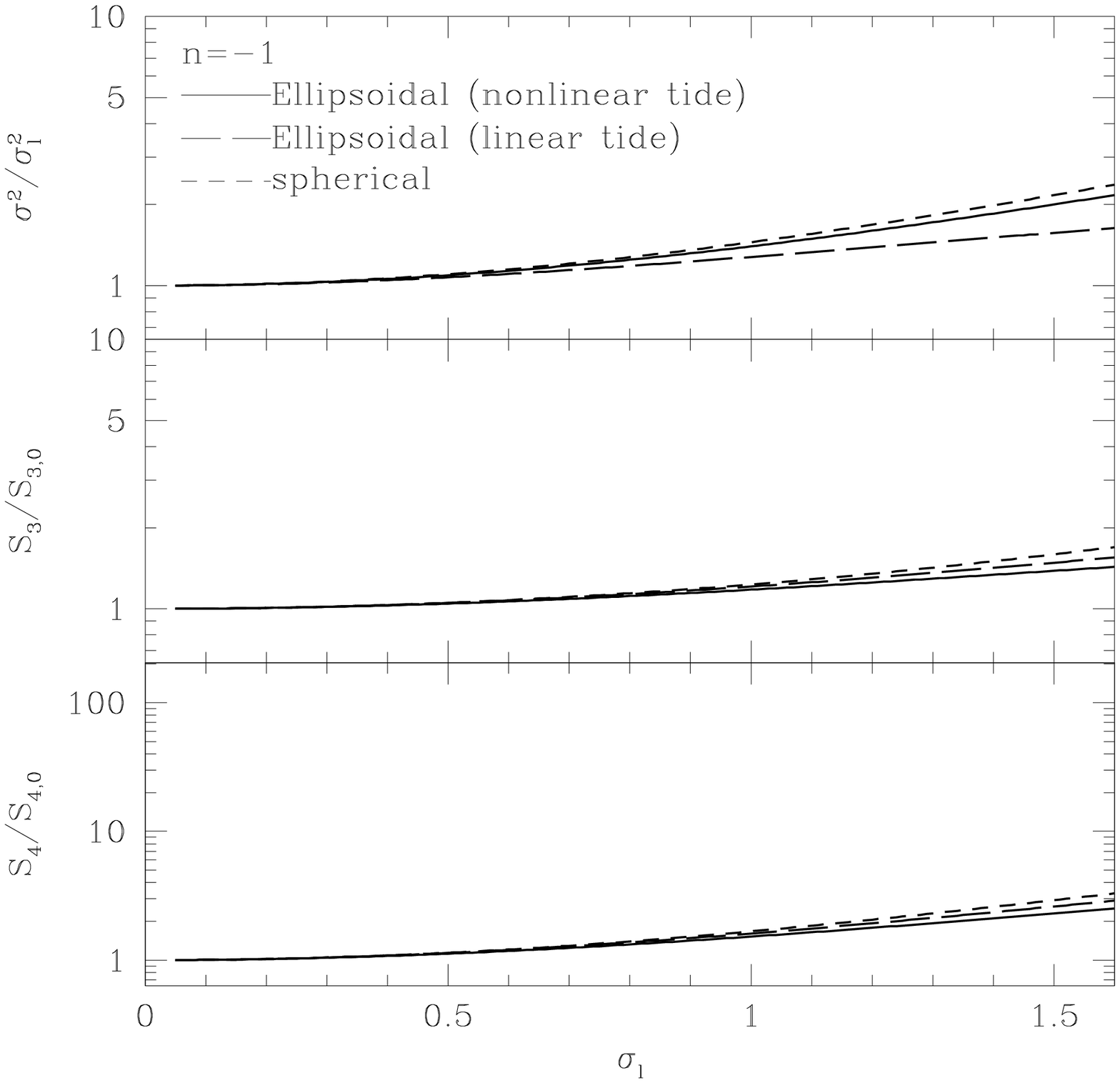}\vspace{0.5cm}\\
 \begin{minipage}{0.49\textwidth}
  \plotone{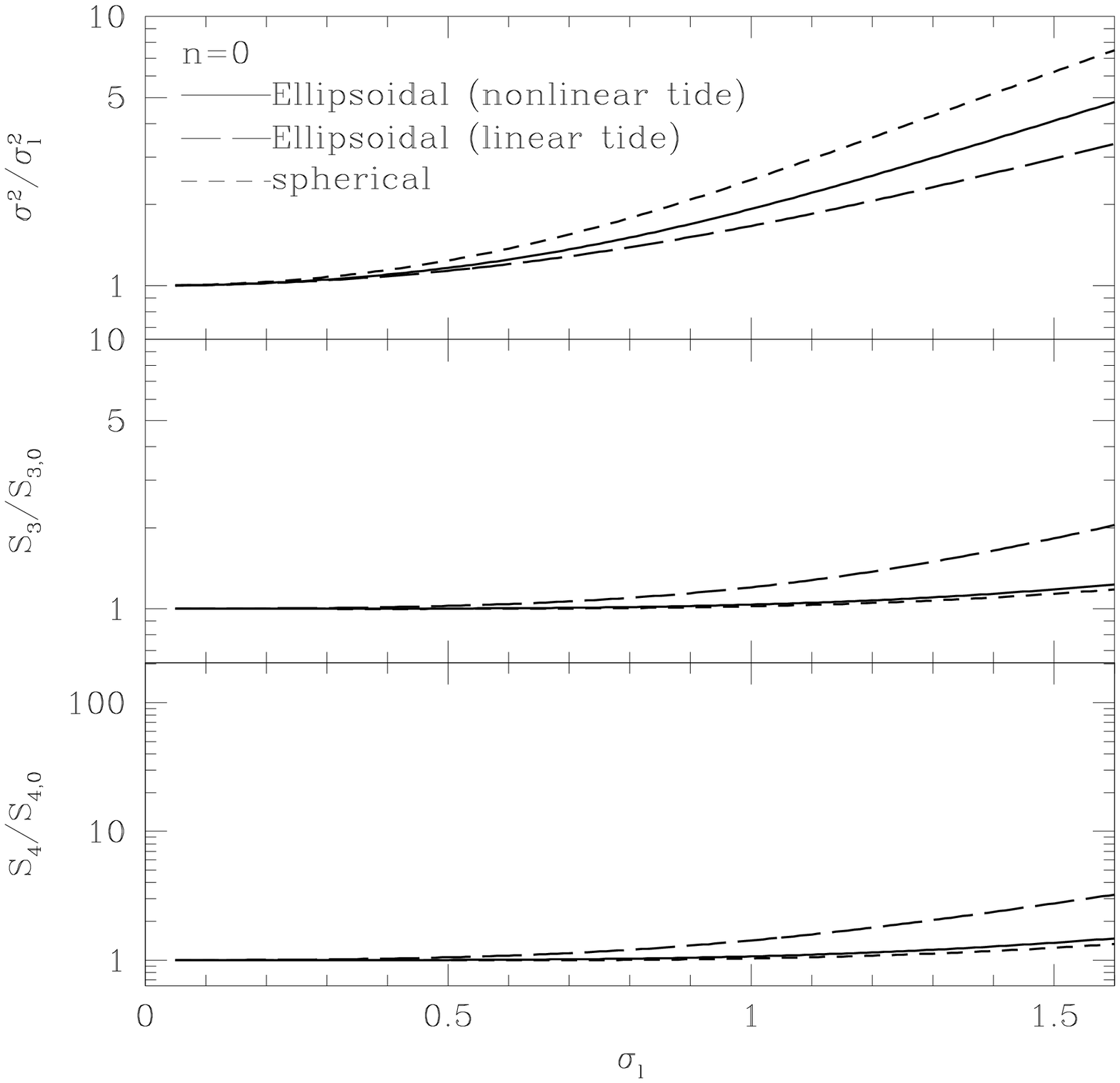}
 \end{minipage} 
 \figcaption{Differences between the model predictions in the variance, 
the skewness and the kurtosis. The    
 perturbative results up to the
 two-loop order are shown in various models of the Lagrangian local dynamics. 
While the {\it solid} ({\it long-dashed}) lines
 represent the predictions based on the ECM with non-linear (linear)
 external tide, the {\it short-dashed} lines indicate the results
 obtained from the SCM \citep{FG1998}.
 \label{fig:ScaleFreeMoment}}
\end{figure}

%%%%%%%%%%%%%%%%%%%%%%%%%%%%%%%%%%%%%%%%%%%%%%%%%%%%%%%%%%%%%%%%%%%%%%
%%%%%%%%%%%%%%%%%%%%%%%%%%%%%%%%%%%%%%%%%%%%%%%%%%%%%%%%%%%%%%%%%%%%%%
%
%
%
%
%
%
To understand both the qualitative and the quantitative behaviors 
of the above two predictions, we here briefly discuss the 
systematic dependence of the perturbative results. 
Table \ref{tbl:coeff} summarizes 
the numerical values of the
coefficients, $s_{2,i}$ up to the three-loop order, and $S_{3,i}$ and
$S_{4,i}$ up to the two-loop order for each model with various spectral
indices. And using these results, 
we plot the cumulants up to the two-loop order  
in figure \ref{fig:ScaleFreeMoment}, where the cumulants are 
normalized by the leading-order results $\sigma_l^2$, $S_{3,0}$ and 
$S_{4,0}$ (from {\it top} to {\it bottom}) and 
are depicted as function of linear variance $\sigma_l$. 
The results from the SCM ({\it short-dashed}) are 
essentially the same results as those obtained by \citet{FG1998}. 
Note that the leading-order results 
for the cumulants in all model predictions  
rigorously coincide with those obtained from 
the exact perturbation theory \citep[e.g.,][]{B1994b}, 
irrespective of the choice of the external tidal term.

Figure \ref{fig:ScaleFreeMoment} shows that 
the differences between the model 
predictions are generally small for the spectral indices $n<0$
and these are expected to become negligible as
decreasing $n$, approaching the non-smoothing results ($n=-3$) as
obtained previously \citep{OKT2003}. For $n=-2$, the predictions up to
the one-loop order give
%%%%%%%%%%%%%%%%%%%%%%%%%%%%%%%%%%%%%%%%%%%%%%%%%%%%%%%%%%%%%%%%%%%%%%
\begin{eqnarray}
 \sigma^2&\approx& \sigma_l^2+0.61\sigma_l^4\\
 S_3&\approx& 3.86+3.21\sigma_l^2
\end{eqnarray}
%%%%%%%%%%%%%%%%%%%%%%%%%%%%%%%%%%%%%%%%%%%%%%%%%%%%%%%%%%%%%%%%%%%%%%
for the SCM \citep{FG1998} and 
%%%%%%%%%%%%%%%%%%%%%%%%%%%%%%%%%%%%%%%%%%%%%%%%%%%%%%%%%%%%%%%%%%%%%% 
\begin{eqnarray}
 \sigma^2&\approx& \sigma_l^2+0.88\sigma_l^4\\
 S_3&\approx& 3.86+3.18\sigma_l^2
\end{eqnarray}
%%%%%%%%%%%%%%%%%%%%%%%%%%%%%%%%%%%%%%%%%%%%%%%%%%%%%%%%%%%%%%%%%%%%%%
for the exact perturbation theory \citep{SF1996b,S1997}.  
Comparing the above results with Table \ref{tbl:coeff}, the numerical
values of the coefficients are close to those from the ECM
prediction. On the other hand, for the $n=0$ case, a large discrepancy
appears in the variance $\sigma^2$ (Fig.\ref{fig:ScaleFreeMoment}). The
skewness and the kurtosis also show
a relatively large difference. These behaviors are indeed consistent
with the PDF shown in figure \ref{fig:PDFscalefreel}. The reason why the
discrepancy in the model predictions become large as increasing $n$ is
partially ascribed to the Jacobian in the smoothed density PDF
(\ref{eq:smoothPDFd}). The expression of the Jacobian
$|\partial\hat{\lambda}_j/\partial\lambda_k'|$ in (\ref{eq:Jacobian})
contains the quantity related to the velocity divergence $\theta$
multiplied by the spectral dependent factor, $\gamma/6$. In previous
study of the non-smoothing case \citep{OKT2003}, we found that while the
differences between the model predictions are almost negligible for the
density fields, a large difference appears in the velocity divergence
$\theta$. This readily implies that the differences between the model
predictions also become significant in the present case, depending on the
factor $\gamma=-(n+3)$. In other words, for a large deviation from
$\gamma=0$, the model predictions sensitively depend on the choice of
the Lagrangian local dynamics in local approximation. That is, not only
the evolution of local density but also the evolution of velocity field
should be devised to approximate the fluid dynamics precisely. This point 
is important and should be kept in mind when comparing the prediction with 
N-body simulations (see Sec.\ref{subsec:scale-free}). 
%
%
%
%
%
%
%
%
%
%
%
%
%%%%%%%%%%%%%%%%%%%%%%%%%%%%%%%%%%%%%%%%%%%%%%%%%%%%%%%%%%%%%%%%%%%%%%
%%%%%%%%%%%%%%%%%%%%%%%%%%%%%%%%%%%%%%%%%%%%%%%%%%%%%%%%%%%%%%%%%%%%%%
\section{Comparison with N-body simulations}
\label{sec:results}
%%%%%%%%%%%%%%%%%%%%%%%%%%%%%%%%%%%%%%%%%%%%%%%%%%%%%%%%%%%%%%%%%%%%%%
%%%%%%%%%%%%%%%%%%%%%%%%%%%%%%%%%%%%%%%%%%%%%%%%%%%%%%%%%%%%%%%%%%%%%%
%
%
%
%
%
%
%
%
%
We are now in position to discuss the validity and the usefulness of the
local approximations comparing the theoretical prediction with N-body
simulations. For this purpose, we specifically use the N-body data for
the scale-free models with initial power spectra $P(k)\propto k^n$
$(n=-2,-1,0)$ \citep{J1998}, as well as the CDM model with cosmological
constant \citep[$\Lambda$CDM;][]{JS1998}. The gravitational force
calculation is based on the P$^3$M algorithm. All the models employ
$N=256^3$ dark matter particles in a periodic comoving cube $\Lbox^3$,
where the box size of the $\Lambda$CDM model is chosen as
$\Lbox=300h^{-1}$Mpc. While the scale-free models assume an Einstein-de
Sitter universe, cosmological parameters of the $\Lambda$CDM model are
set as $(\Omega_{\rm m},\Omega_\Lambda,h,\sigma_8)=(0.3,0.7,0.7,1.0)$,
where the normalization $\sigma_8$ means the linear rms fluctuation at
top-hat smoothing radius $R=8h^{-1}$Mpc. For the scale-free models, the
normalization of the density fluctuation is determined by setting the
linear rms fluctuation to unity at $R=0.1\Lbox$. Below, we first present
the results in $\Lambda$CDM model (Sec.\ref{subsec: LCDM}). The PDFs in
the scale-free models are compared in section
\ref{subsec:scale-free}. In comparing the prediction with simulation, 
we also present the perturbation results of local approximation obtained in 
previous section. 
%
%
%
%
%
%%%%%%%%%%%%%%%%%%%%%%%%%%%%%%%%%%%%%%%%%%%%%%%%%%%%%%%%%%%%%%%%%%%%%%
\subsection{$\Lambda$CDM model}
\label{subsec: LCDM}
%%%%%%%%%%%%%%%%%%%%%%%%%%%%%%%%%%%%%%%%%%%%%%%%%%%%%%%%%%%%%%%%%%%%%%
%
%
%
%
%
%
%
%
%
%
The validity of the local approximation using the SCM has been
previously studied by \citet{SG2001} in the case of the standard CDM
model and a good agreement with N-body simulation was found. We thus
expect that the local approximation with both the SCM and the ECM also
provides an excellent agreement with N-body simulation in the case of
the $\Lambda$CDM model.

\begin{figure}
 \plotone{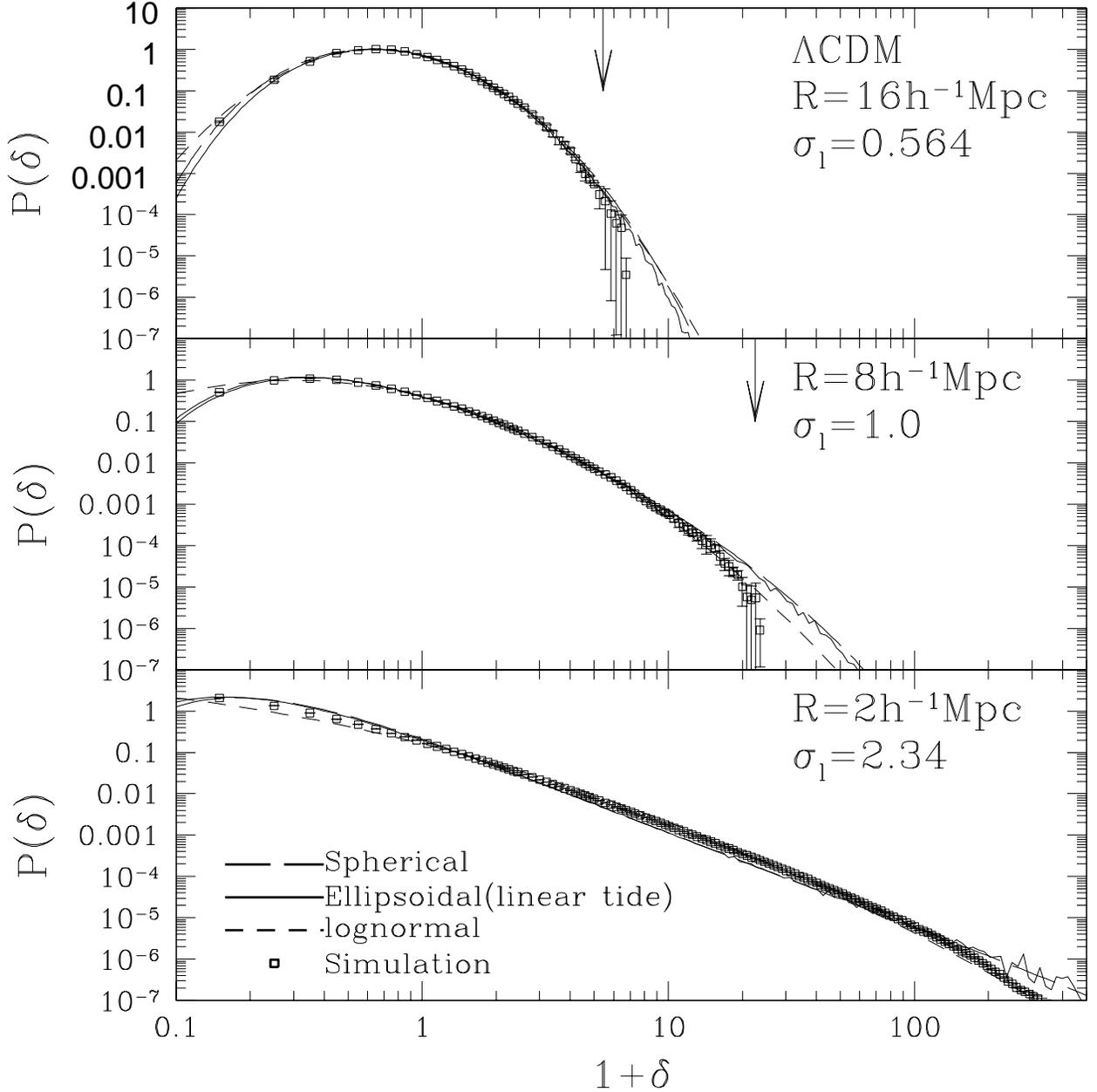}
 \figcaption{PDFs of density field in $\Lambda$CDM model with top-hat
 smoothing window; $R=2$({\it bottom}), $8$({\it middle}) and
 $16h^{-1}$Mpc({\it top}). The {\it open-squares} represent the N-body
 results. The error bars indicate the $1\sigma$ variation among three
 different realizations. Values of $\sigma_l$ in each panel are the
 linear variance at each smoothing radius. The arrows indicate 
 the mean value of the cutoff density $\delta_{\rm max}$. Note that 
 the cutoff density at $R=2h^{-1}$Mpc reaches at $601$. 
{\it Solid} lines: the prediction based on the ECM with linear external 
tide. {\it Long-dashed} lines: the prediction obtained from the SCM.
 {\it Short-dashed} lines: the lognormal PDF adopting the variance
 $\sigma^2$ calculated directly from the simulations.
 \label{fig:PDFcdml}}
\end{figure}

Figure \ref{fig:PDFcdml} shows the PDFs obtained from the N-body data
for the top-hat smoothed density fields ({\it open-squares}). The error
bars indicate the $1\sigma$ errors among three different
realizations. The smoothing radii are chosen as $R=16$, $8$,
$2h^{-1}$Mpc (from {\it top} to {\it bottom}). In figure
\ref{fig:PDFcdml}, the PDFs from the SCM and the ECM with linear
external tide are depicted as {\it long-dashed} and {\it solid} lines,
respectively. We also calculated the PDFs from the ECM with non-linear
external tide, but the results are almost the same as obtained from the
ECM with linear external tide.
Clearly, these models almost coincide with each other and
the agreement with N-body results is excellently good. For comparison,
we also plot the empirical model of the lognormal distribution ({\it
short-dashed}):
%%%%%%%%%%%%%%%%%%%%%%%%%%%%%%%%%%%%%%%%%%%%%%%%%%%%%%%%%%%%%%%%%%%%%%
\begin{equation}
 P(\delta)=\frac{1}{\sqrt{2\pi}\sigma_{\rm LN}}\frac{1}{1+\delta}\exp\left[-\frac{1}{2\sigma_{\rm LN}^2}\left(\log(1+\delta)+\frac{\sigma_{\rm LN}^2}{2}\right)^2\right]
  \label{eq:lognormal_PDF}
\end{equation}
%%%%%%%%%%%%%%%%%%%%%%%%%%%%%%%%%%%%%%%%%%%%%%%%%%%%%%%%%%%%%%%%%%%%%%
with $\sigma_{\rm LN}^2=\log(1+\sigma^2)$. Here the
quantity $\sigma$ denotes the variance of the local density field, which
is estimated from the N-body simulation. Albeit the simplicity of the
analytical expression (\ref{eq:lognormal_PDF}), the lognormal PDFs also
approximate the N-body results quite accurately. Agreement with N-body
simulation still remains good even at the high-density tails of small
radii, $R=2$ and $8h^{-1}$Mpc. \citet{BK1995} discuss the successful 
lognormal fit of the PDF in the CDM models based on the perturbation results.

\begin{figure}
  \plottwo{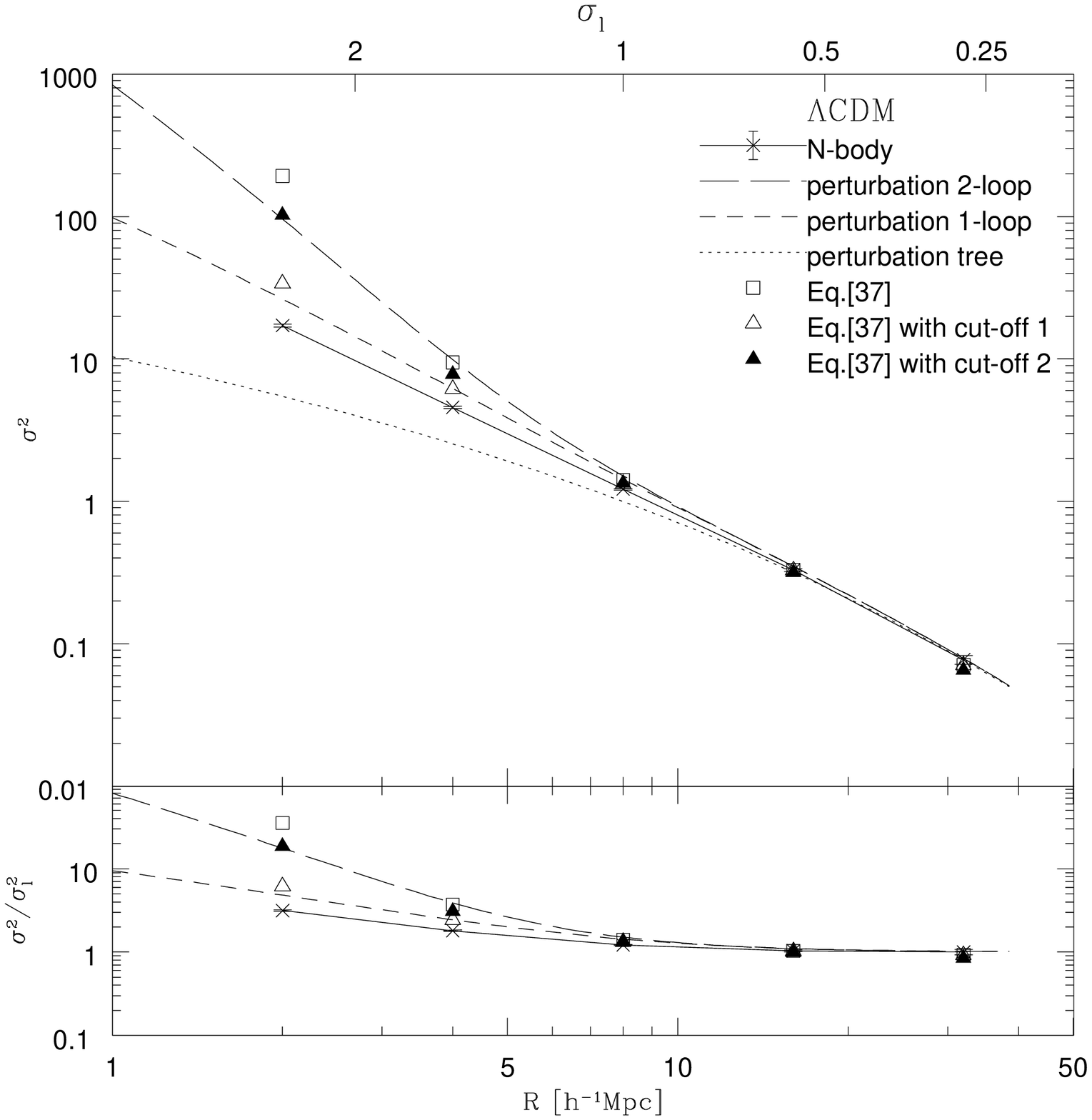}{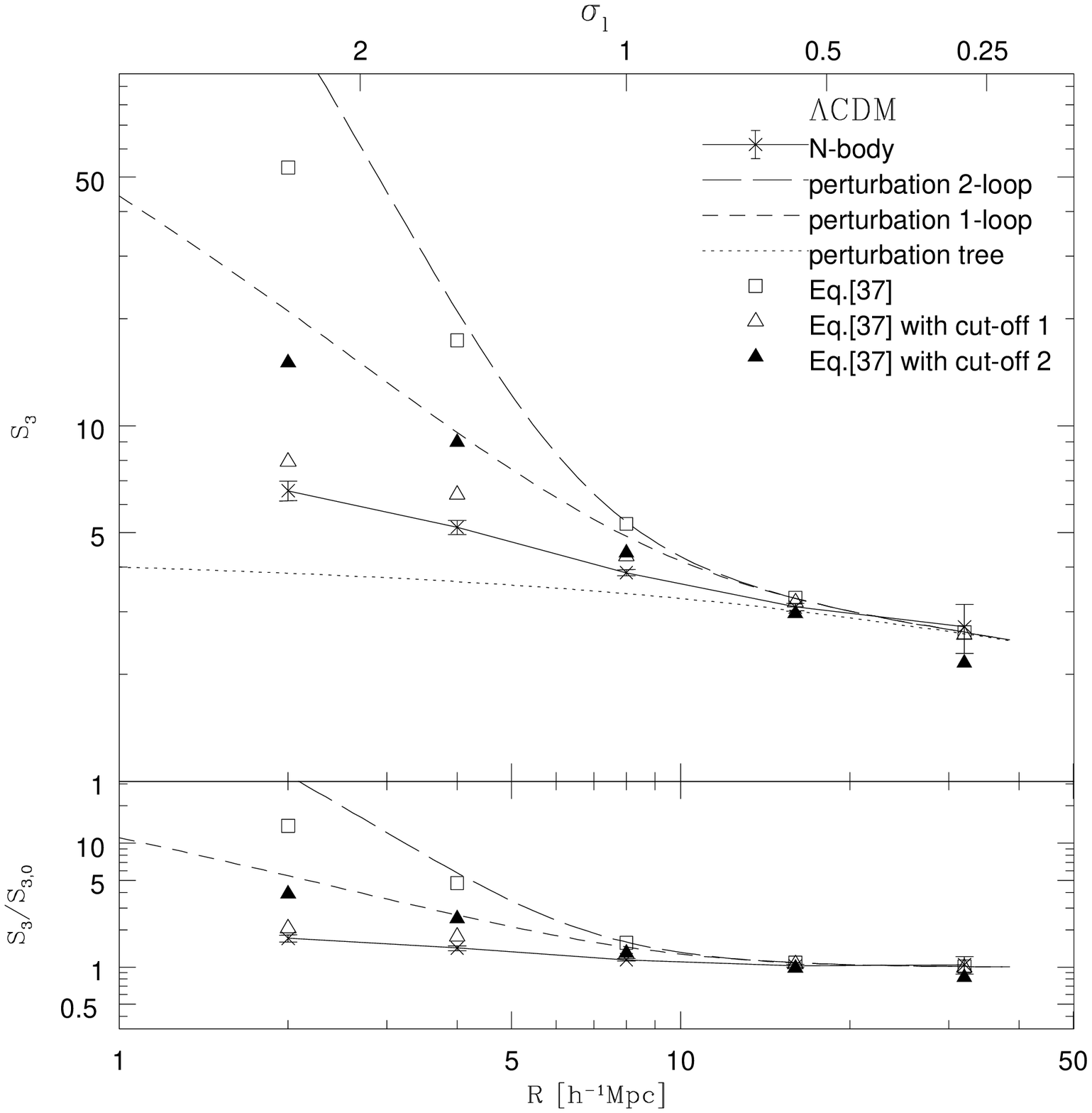}\vspace{0.5cm}\\
 \begin{minipage}{0.49\textwidth}
  \plotone{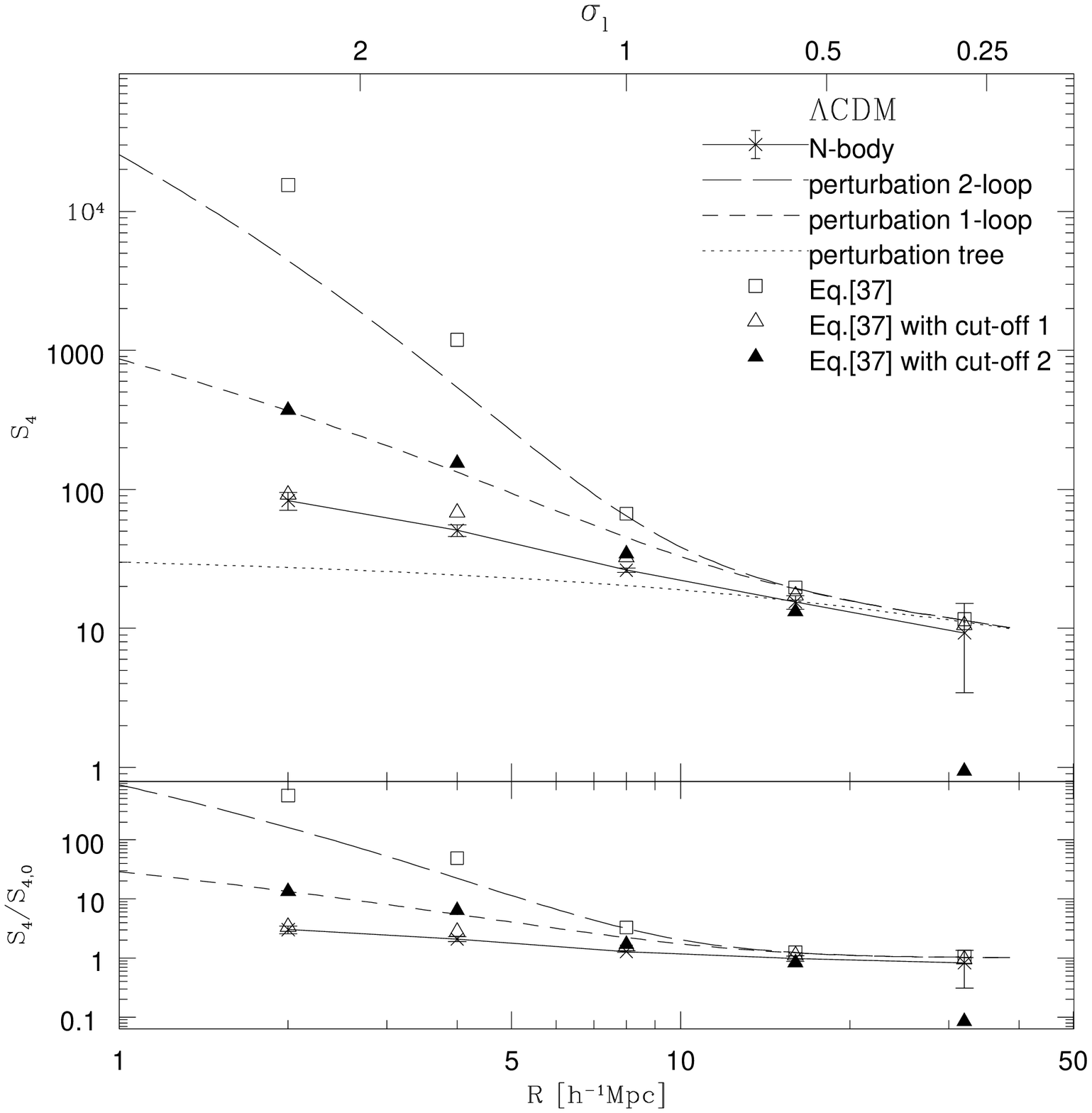}
 \end{minipage} 
 \figcaption{Variance ({\it top}), skewness ({\it middle}) and
 kurtosis ({\it bottom}) of the density field in $\Lambda$CDM model as
 function of smoothing radius $R$. The {\it crosses} with error bars
 represent the results from N-body simulations. The {\it open-squares}
 show the prediction from the ECM with linear external tide based on a
 full knowledge of the PDF (eq.[\ref{eq:MomentFromPdf}]). The
 {\it open-} and {\it filled-triangles} are the same prediction 
as {\it open-squares}, but taking account of the limited range of the 
density PDF, $[\delta_{\rm min},\delta_{\rm max}]$ (cutoff-1, cutoff-2). 
While the cutoff values 
in  {\it open-squares} are estimated from the N-body data, 
$\delta_{\rm min}$ and $\delta_{\rm max}$ in 
{\it filled-triangles} are determined from equation (\ref{eq:deltamax_min}) 
with a help of theoretical PDF. {\it Long-dashed} lines: the
 perturbative predictions of cumulants based on the ECM up to the
 two-loop order. {\it Short-dashed} lines: the perturbative predictions up to
 the one-loop order. {\it Dotted} lines: the leading order results of
 the perturbation theory.
 \label{fig:cdmmoment}}
\end{figure}

In order to check the accuracy of the model predictions, we quantify the
cumulants of the density fields. In figure \ref{fig:cdmmoment}, we plot
the variance $\sigma^2$, the skewness
$S_3\equiv\langle\delta^3\rangle/\sigma^4$ and the kurtosis
$S_4\equiv(\langle\delta^4\rangle-3\sigma^4)/\sigma^6$ as a function of
smoothing radius (from {\it top} to {\it bottom}). In each panel, the
{\it crosses} with error bars indicate the results obtained from the
N-body simulations, while the {\it open-squares} show the results from
the local approximation with ECM, in which the moments
$\langle\delta^N\rangle$ are calculated from the full knowledge of the
PDF $P(\delta)$ (see eq.[\ref{eq:MomentFromPdf}]). On the other hand,
the short-dashed and the long-dashed lines are the perturbative calculations
of the cumulants based on the ECM up to the one-loop order and the
two-loop order, respectively(see Sec.\ref{subsubsec:linear-tide}). 
As a reference, we also plot the leading-order(tree-level) prediction in 
dotted lines.  
In contrast to a naive expectation from figure \ref{fig:PDFcdml}, 
the ECM prediction based on the 
PDF significantly deviates from the N-body results at the smaller
radius, $R\simlt 8h^{-1}$Mpc, although it roughly match the perturbation
results up to the two-loop order.

\begin{figure}
 \plotone{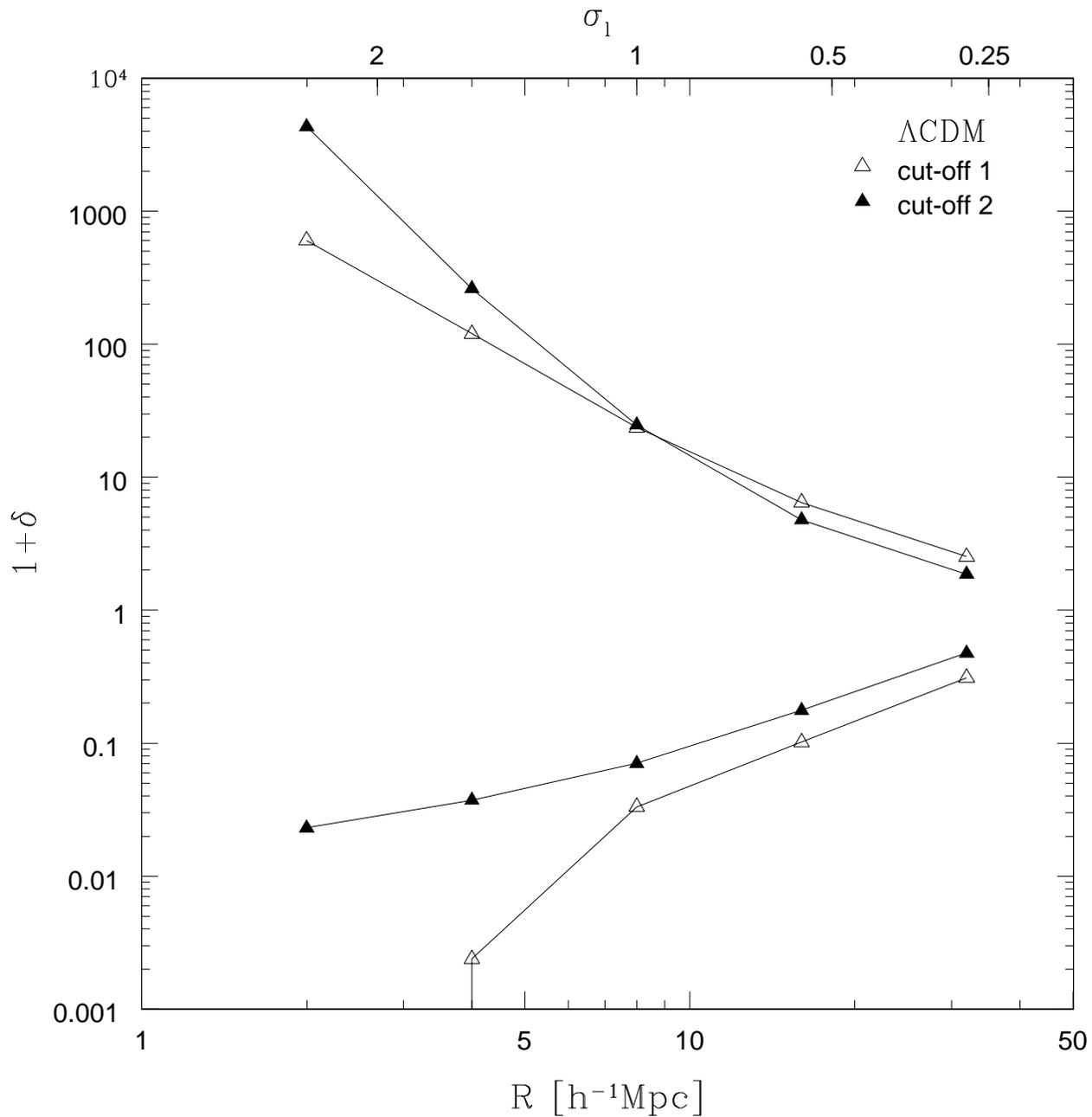}
 \figcaption{The cutoff values of the density field $\delta_{\rm min}$ and 
$\delta_{\rm max}$ as function of smoothing radii. While the 
{\it open-triangles} show the cutoff density estimated from the N-body 
simulations, the 
{\it filled-triangles} represent the values obtained from the finite sampling 
effect (\ref{eq:deltamax_min}). 
 \label{fig:deltaminmax}}
\end{figure}

\citet{KTS2001} remarked that the origin of this discrepancy 
might be due to the fact that the density field $\delta$ in N-body
simulation does not extend the entire range between $-1$ and $+\infty$,
but rather is limited in the range,
$\delta_{\rm min}<\delta<\delta_{\rm max}$ owing to the finite size of
the simulation box. Indeed, a closer look at figure \ref{fig:PDFcdml} 
reveals that there exist the sharp cutoff at the high-density tails of 
simulated PDF({\it arrows} in each panel). 
To examine the significance of this effect, 
the ECM predictions of the cumulants taking account of the finite range 
$[\delta_{\rm min}, \delta_{\rm max}]$ are plotted in figure
\ref{fig:cdmmoment} (cutoff-1, {\it open-triangles}). The cutoff values
$\delta_{\rm min}$ and $\delta_{\rm max}$ are estimated from the N-body
data. Further, in figure \ref{fig:cdmmoment}, we plot the prediction taking 
account of the cutoff values determined from the simple 
assumptions that (i) the major effect for the finite range of $\delta$ 
comes from the finite sampling effect and (ii) the theoretical PDF 
is correct if the box size of the simulation becomes infinite. 
We then have \citep{KTS2001}:
%%%%%%%%%%%%%%%%%%%%%%%%%%%%%%%%%%%%%%%%%%%%%%%%%%%%%%%%%%%%%%%%%%%%%%%
\begin{equation}
 \frac{\Lbox^3}{4\pi R^3/3}
\int^{\delta_{\rm min}}_{-1}P(\delta)d\delta=1,~~~~~~
 \frac{\Lbox^3}{4\pi R^3/3}
\int_{\delta_{\rm max}}^{\infty}P(\delta)d\delta=1,  
\label{eq:deltamax_min}
\end{equation}
%%%%%%%%%%%%%%%%%%%%%%%%%%%%%%%%%%%%%%%%%%%%%%%%%%%%%%%%%%%%%%%%%%%%%%%
where $P(\delta)$ is the PDF obtained from the ECM. 
The resultant cutoff values $(\delta_{\rm min}, \delta_{\rm max})$ are 
plotted as function of 
smoothing radii in figure \ref{fig:deltaminmax}.

The resultant amplitudes of the cumulants are significantly reduced  
and the model prediction turns out to reproduce the simulation data very
well in the case using the N-body data (cutoff-1). 
This readily implies that the apparent discrepancy in the
cumulants mainly comes from the limited range of the density in PDF and
one concludes that the local approximation with ECM provides an accurate
prediction for both the density PDF and the cumulants in the
$\Lambda$CDM model. Similarly, one expects that the accuracy of the
model prediction still remains good for the local approximation with
SCM. Note, however, that the predictions  
with (\ref{eq:deltamax_min}) still exhibits the discrepancy at the 
smaller scales $R\simlt4h^{-1}$Mpc. 
This means that the finite sampling effect might be a major numerical 
effect for the limited range of the density, but still not enough. 
The discreteness effect could be an important source for causing the 
discrepancy on small scales. This point is particularly important for 
self-consistent calculation of cumulants and 
should be treated carefully. Keeping these remarks in mind, 
we next proceed to the scale-free models. 
%
%
%
%
%
%
%
%
%%%%%%%%%%%%%%%%%%%%%%%%%%%%%%%%%%%%%%%%%%%%%%%%%%%%%%%%%%%%%%%%%%%%%%
\subsection{Scale-free models}
\label{subsec:scale-free}
\begin{figure}
  \plottwo{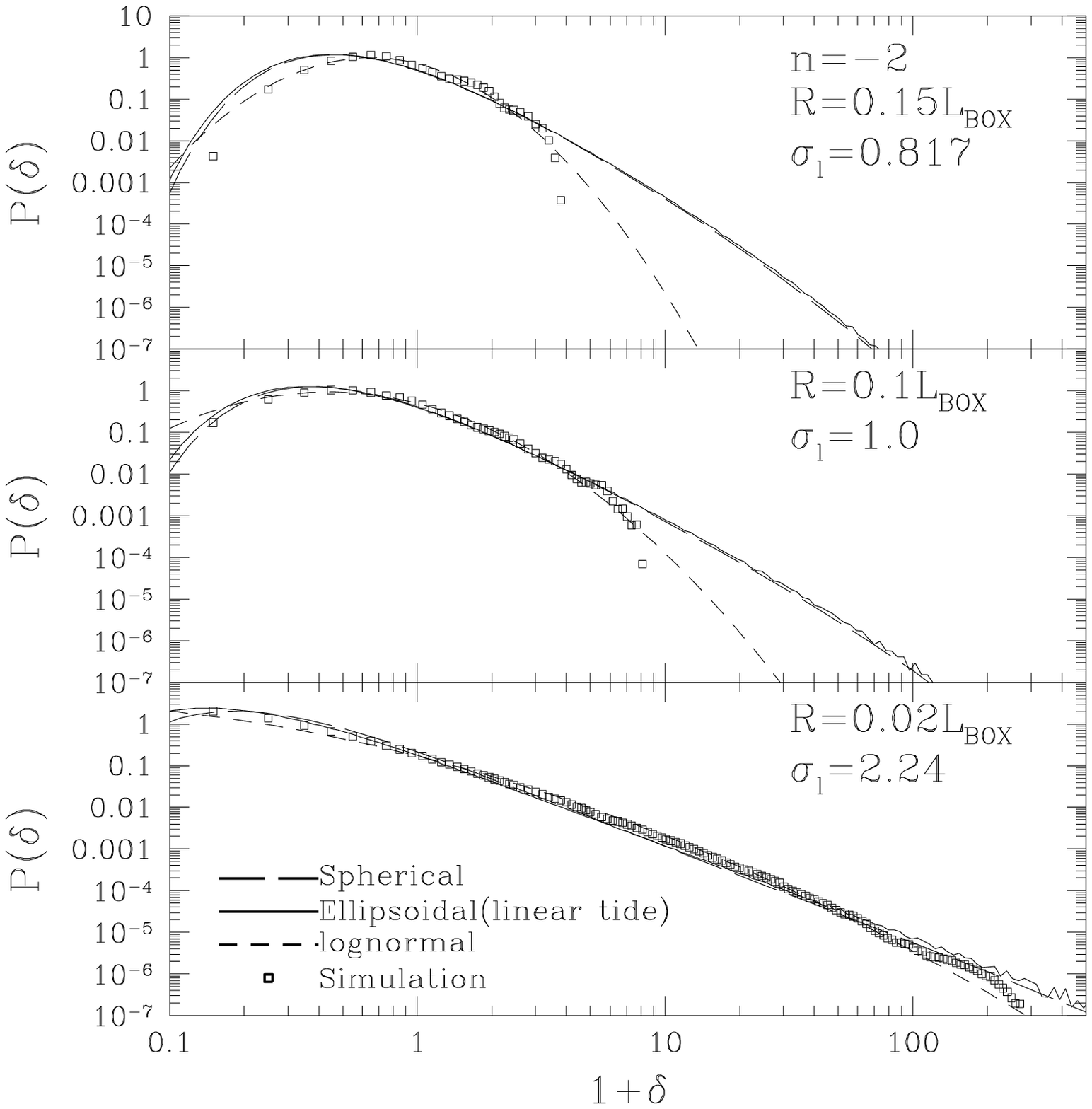}{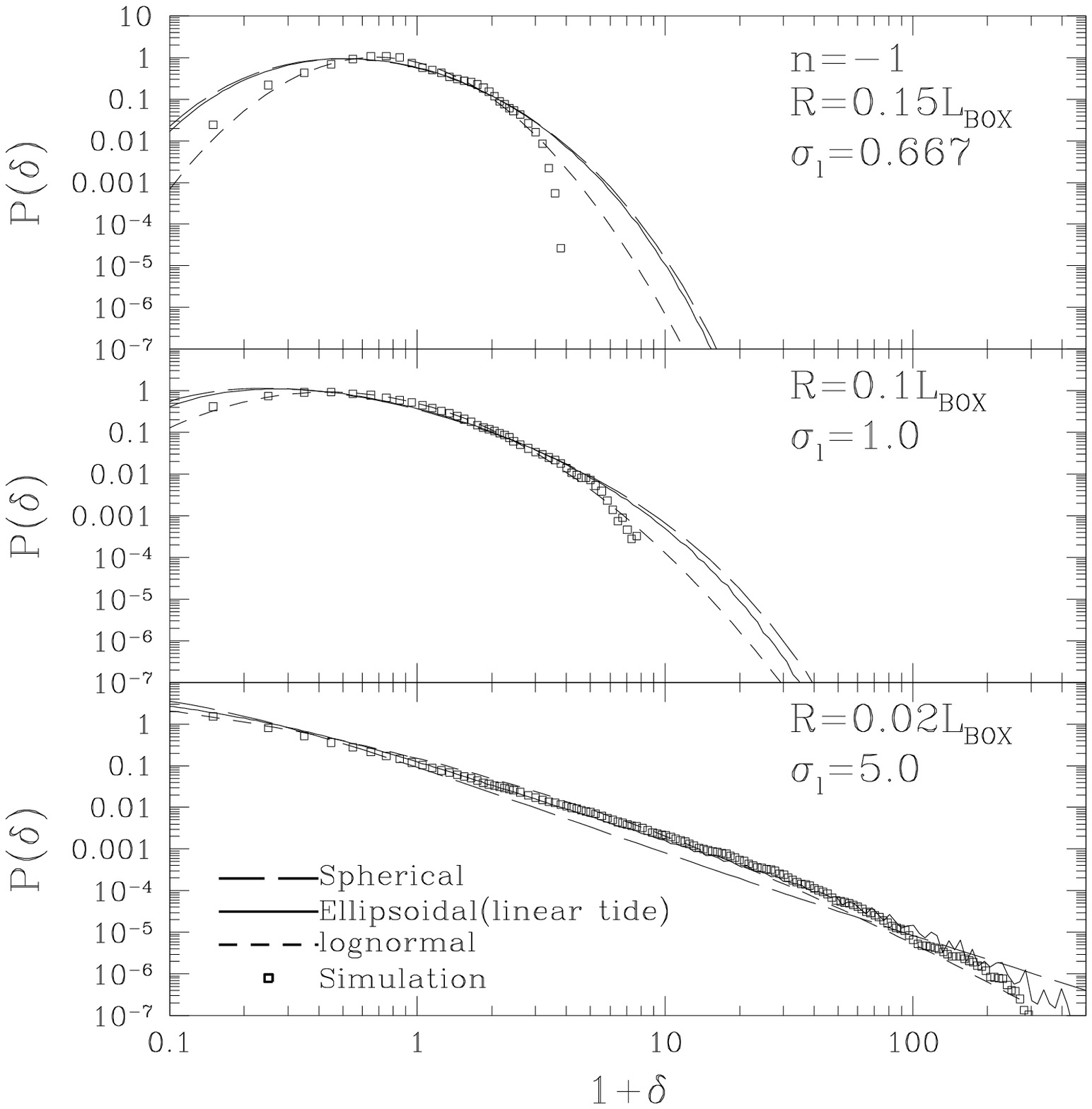}\vspace{0.5cm}\\
 \begin{minipage}{0.49\textwidth}
  \plotone{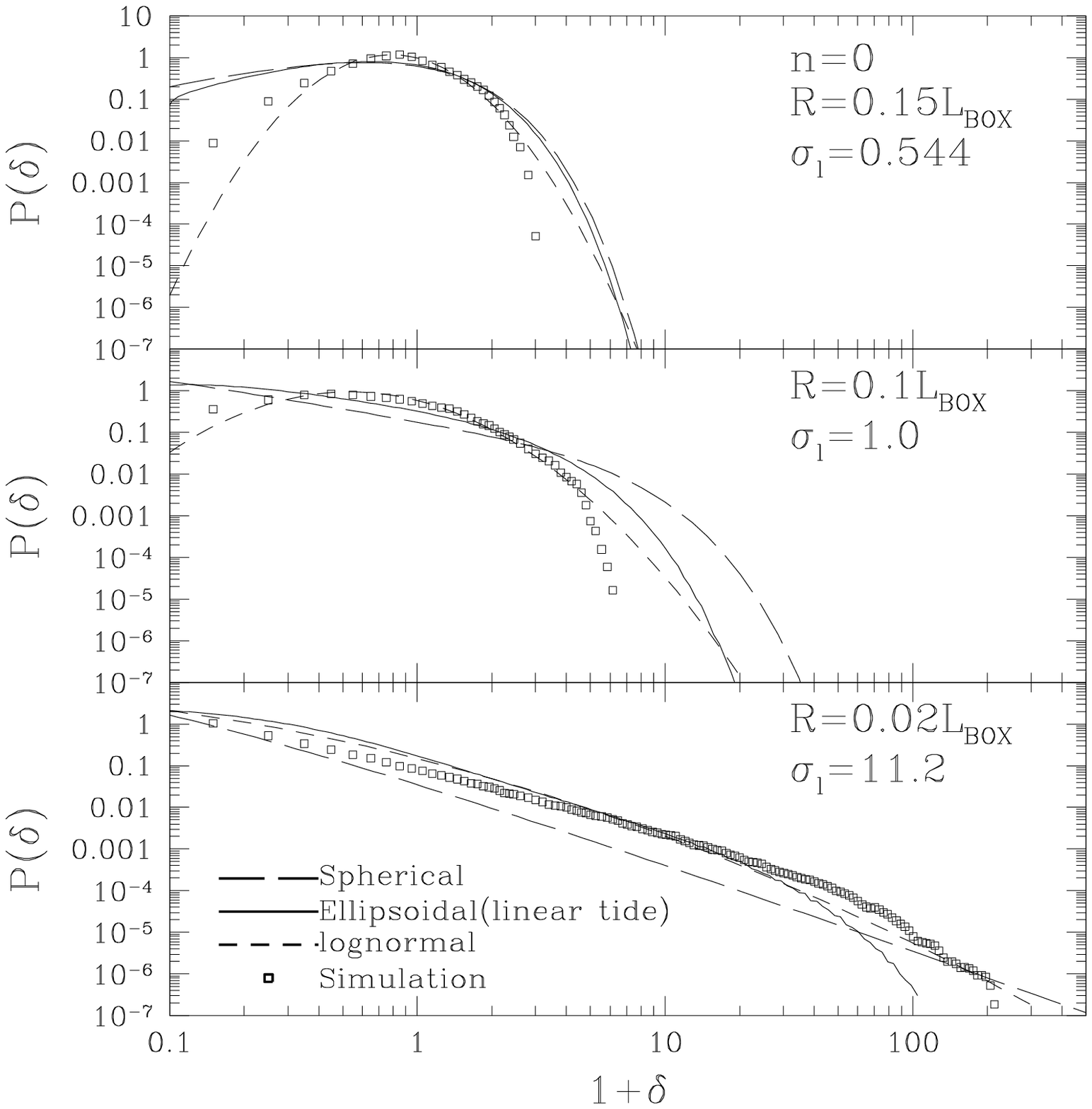}
 \end{minipage} 
 \figcaption{Same as Fig.\ref{fig:PDFcdml}, but in the scale-free
 models ($n=-2,-1,0$);   
 $R=0.02$ ({\it bottom}), $0.05$ ({\it middle}) and
 $0.15\Lbox$ ({\it top}).
 \label{fig:PDFscalefreel}}
\end{figure}

\begin{figure}[hbtp]
  \plottwo{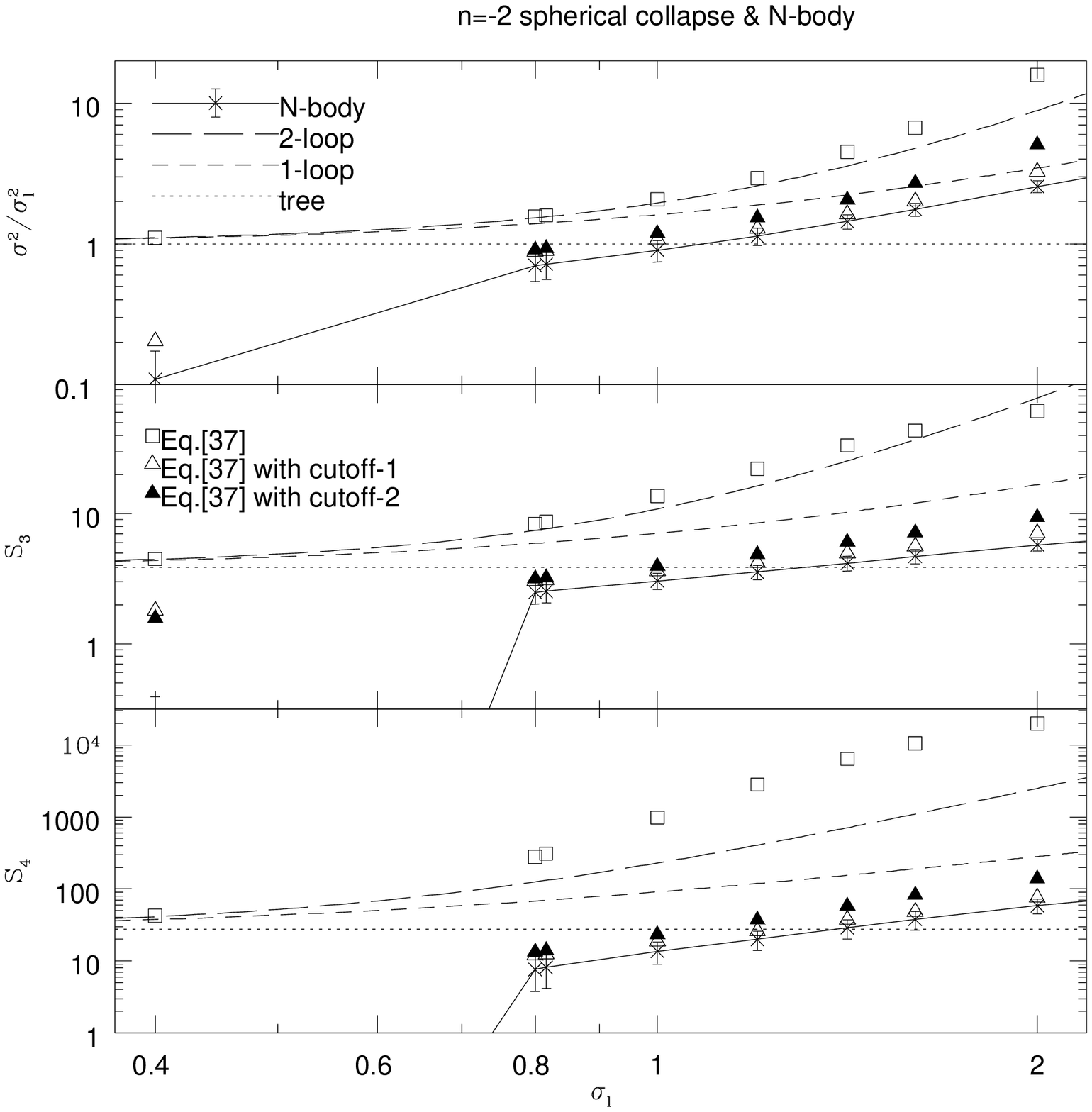}{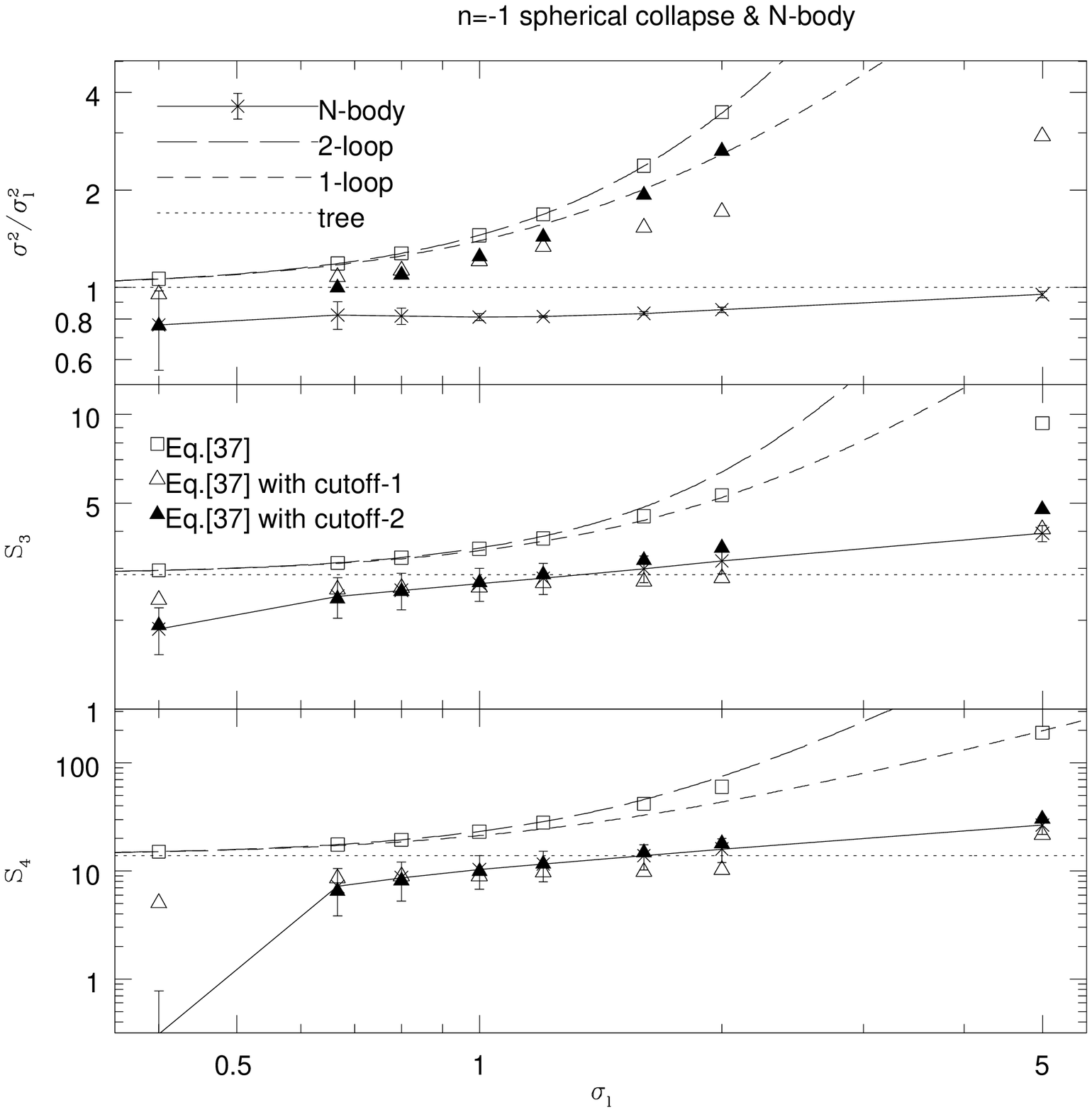}\vspace{0.5cm}\\
 \begin{minipage}{0.49\textwidth}
  \plotone{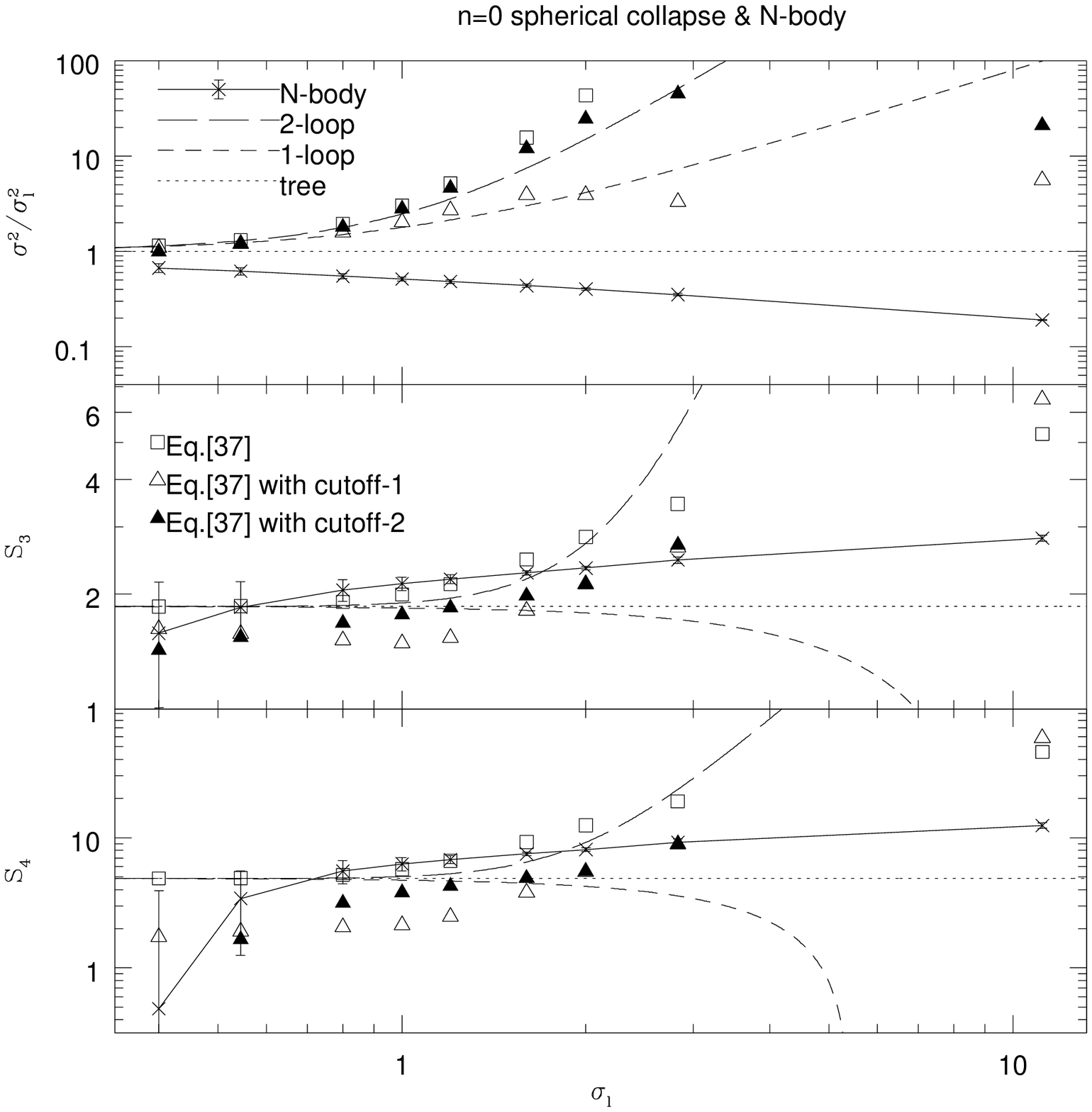}
 \end{minipage} 
 \figcaption{The variance, the skewness and the kurtosis of the N-body
 results in scale-free models, compared with the SCM predictions. 
The {\it crosses} with error bars and {\it solid} line
 represent the N-body simulations. The {\it squares}
 show the predictions from the SCM based on a full knowledge of the PDF
 (eq.[37]). The {\it open-, filled-triangles} are the same as {\it squares},
 but we take account of the cutoff of the density field (cutoff-1, cutoff-2). 
 While the {\it long-dashed} 
 lines represent the perturbative predictions up to the two-loop
 order,  the {\it short-dashed} lines indicate the results up to the
 one-loop order. As a reference, we also plot the leading-order results of 
 perturbation theory in {\it dotted} line.
  \label{fig:cumulants_scalefree_SCM}}
\end{figure}

\begin{figure}[hbtp]
  \plottwo{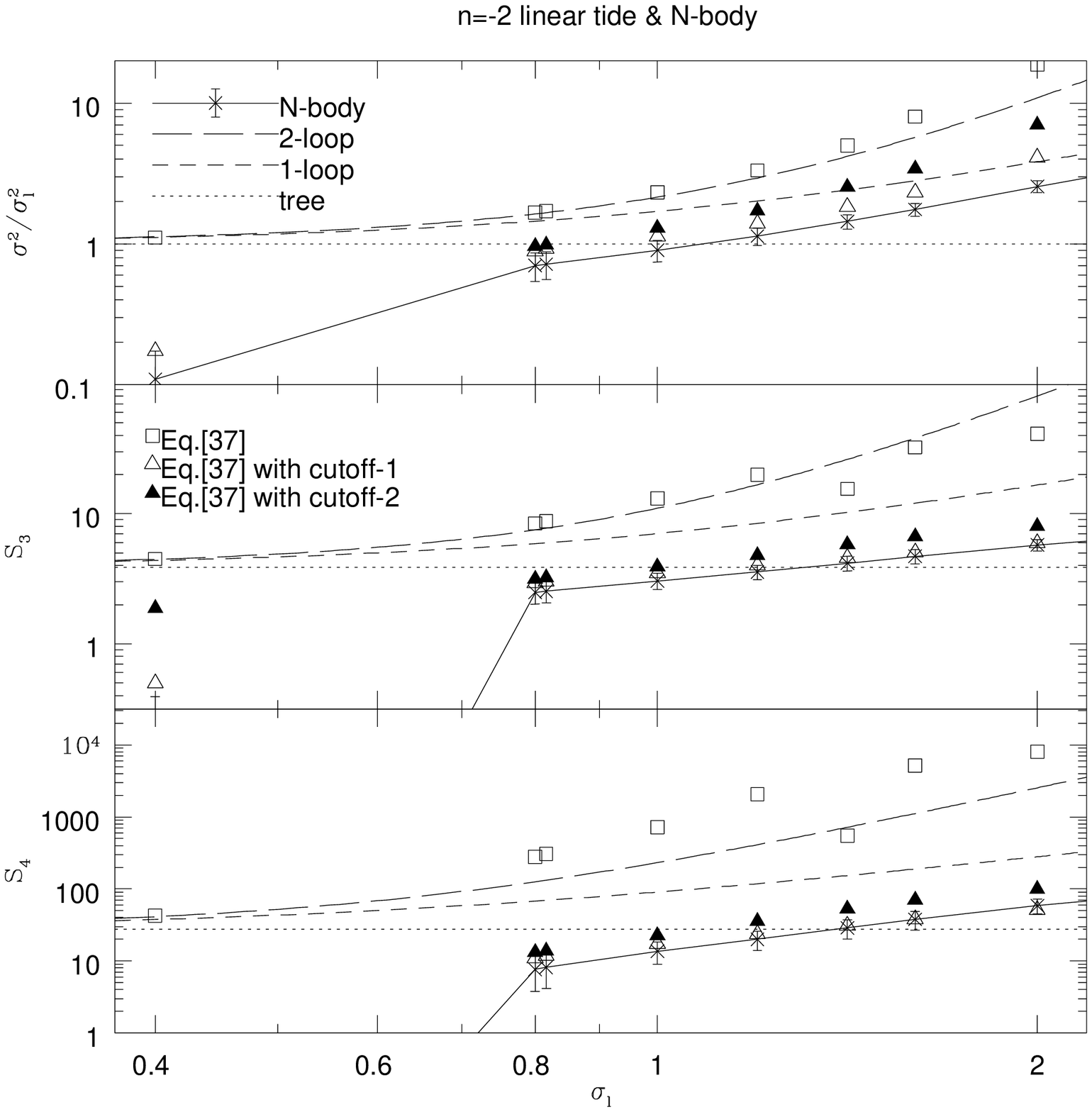}{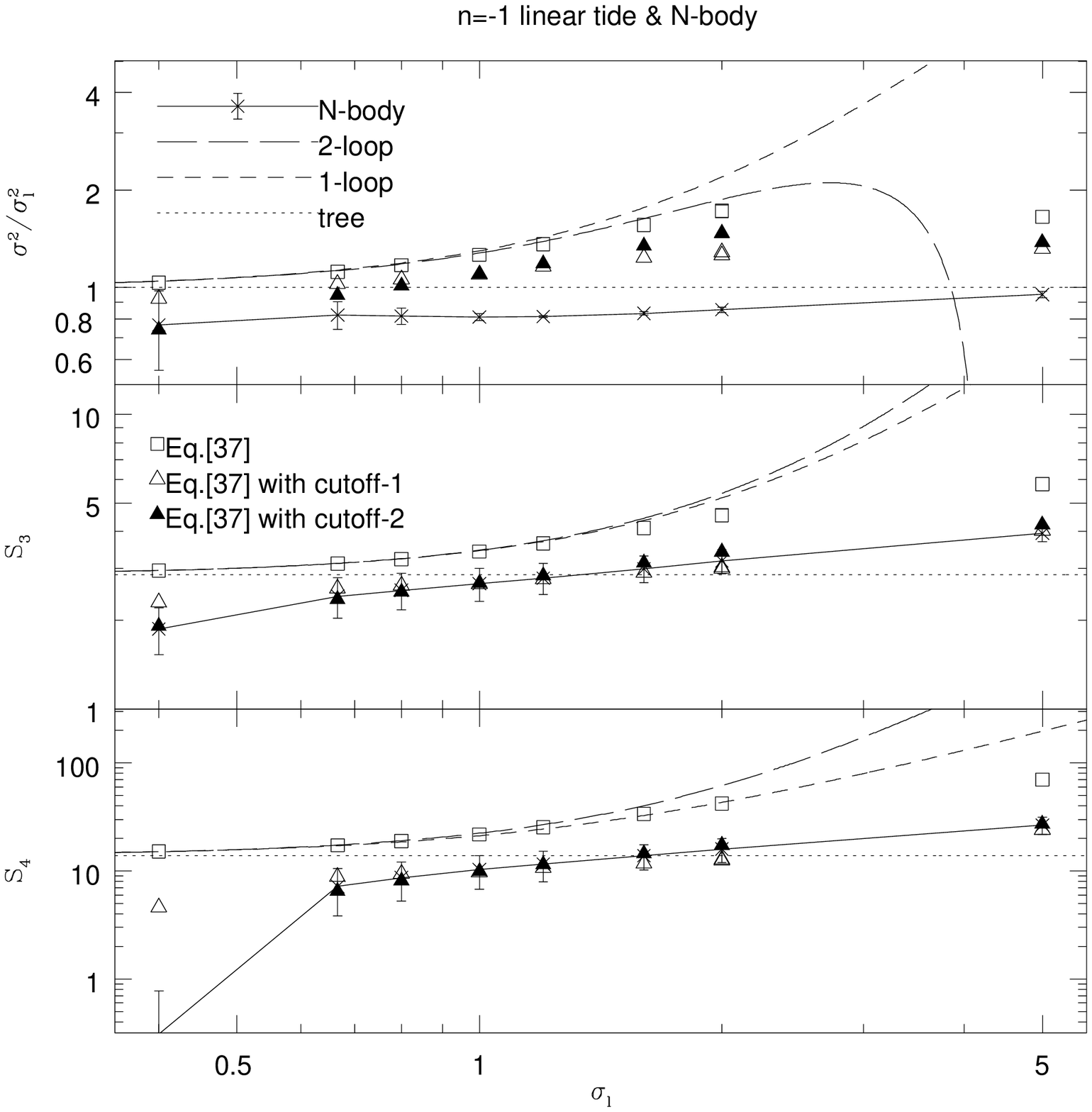}\vspace{0.5cm}\\
 \begin{minipage}{0.49\textwidth}
  \plotone{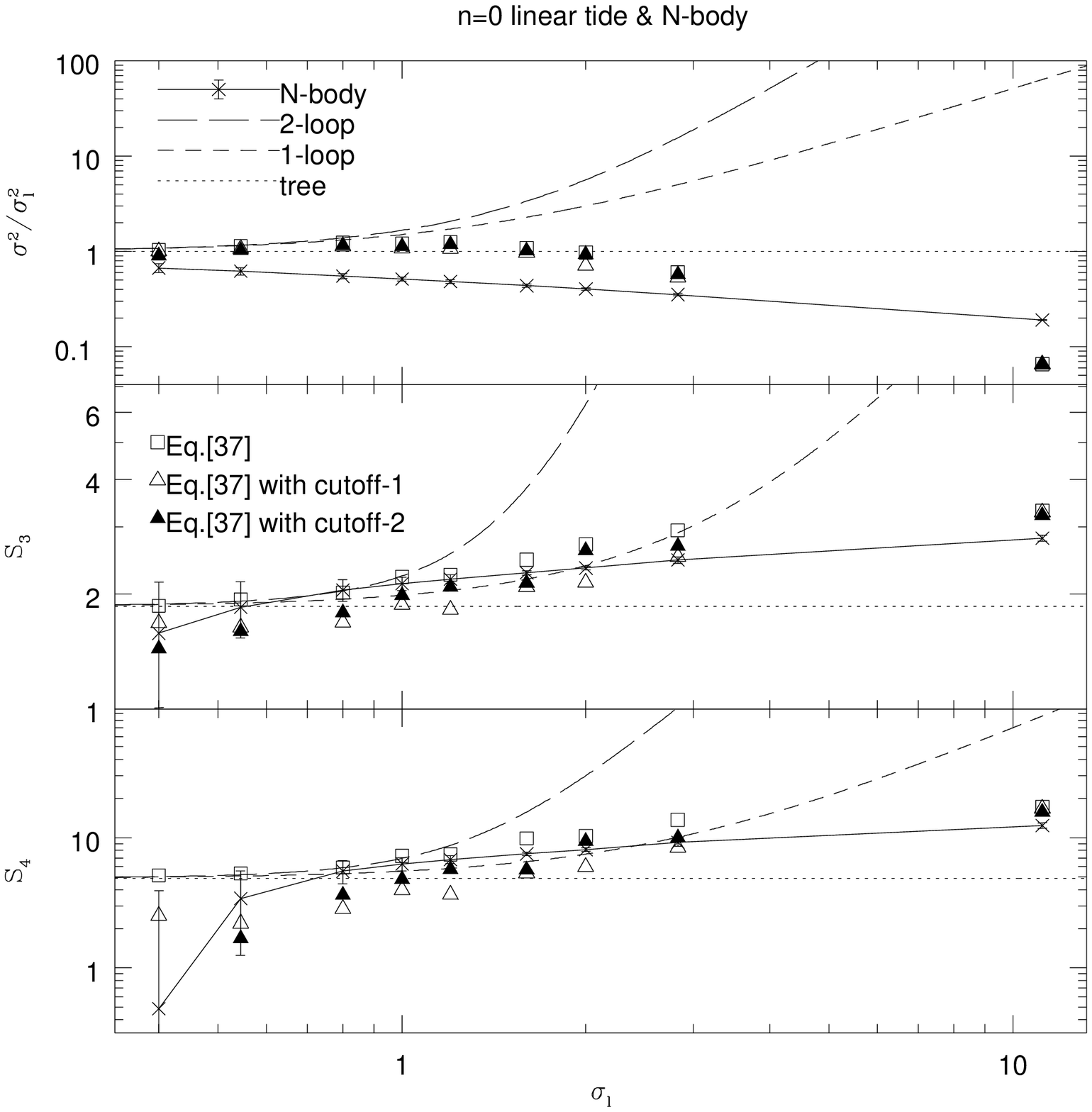}
 \end{minipage} 
 \figcaption{Same as figure \ref{fig:cumulants_scalefree_SCM}, 
but the theoretical predictions based on
 the ECM with linear external tide approximation are presented. 
\label{fig:cumulants_scalefree_ECM}}
\end{figure}

%%%%%%%%%%%%%%%%%%%%%%%%%%%%%%%%%%%%%%%%%%%%%%%%%%%%%%%%%%%%%%%%%%%%%%
%
%
%
%
%
%
%
%
N-body simulations in scale-free model with 
index $n\leq-1$ generally suffer from 
spurious numerical effects compared to the CDM case, 
which significantly affects the statistical 
properties of mass distribution \citep[e.g.,][]{CBH1996}. 
This is not exceptional in our case. Figure \ref{fig:PDFscalefreel} shows 
the density PDFs for $n=-2$ to $0$ models at different smoothing 
radii; $R=0.15$, $0.1$ and $0.02\Lbox$. For larger smoothing radii, 
the PDF obtained from the simulations ({\it open-squares}) has a sharp 
cutoff at high-density tails. Because of this, the amplitude of the cumulants 
is significantly reduced and the N-body simulation fails to even reproduce 
the leading-order results of perturbation theory (see 
Figs.\ref{fig:cumulants_scalefree_SCM} and \ref{fig:cumulants_scalefree_ECM}). 
Nevertheless, focusing on the moderately non-Gaussian tails of PDF, 
we find that the predictions based on the SCM and ECM shows a 
good agreement with N-body simulations for smaller spectral indices, 
$n=-2,\,-1$. Especially at the non-linear scale ($R=0.02\Lbox$), 
the PDF from the ECM seems to improve the prediction, compared to 
the results obtained from the SCM. This is even true for the spectral 
index $n=0$, indicating that the ECM provide a more physical model of
Lagrangian
local dynamics than the SCM. Interestingly, the lognormal model also provides 
a good approximation 
to the N-body simulations, irrespective of the nature of the initial spectra 
and the smoothing scales. Apparently, this contradicts 
with the perturbation results which predict the strong spectral 
dependence in weakly non-linear regime \citep[]{B1994b}. As stated by 
\citet{B1994a} and \citet{BK1995}, however, the perturbation results themselves
resemble the lognormal PDF near the index $n=-1$. Further, 
a systematic comparison with lognormal prediction done by \citet{KTS2001}, 
who basically used the same N-body data as ours, shows that the lognormal model
prediction tends to deviate from simulation for $n=0,~+1$ in 
the weakly non-linear regime even if the cutoff of the PDF is taken into 
account. Thus, no serious contradiction is found at least at the quality
of our data.

Now, consider the cumulants of the density fields. Figures  
\ref{fig:cumulants_scalefree_SCM} and \ref{fig:cumulants_scalefree_ECM} 
show the variance, the skewness and the kurtosis, which are 
compared with the prediction from the SCM and the ECM, respectively.  
As anticipated from figure \ref{fig:PDFscalefreel}, the amplitude of the 
cumulants from N-body data are significantly reduced and the prediction 
from local approximation without cutoff ({\it open-squares}) generally 
overpredicts the simulation results. A more serious aspect is that the 
simulation does not converge to the tree-level results of 
perturbation theory. Even the predictions up to the one-loop correction 
overpredict the simulation results.  As previously remarked by
\citet{CBH1996},
the recovery of tree-level results is difficult for the limited size of 
N-body simulations and a care 
must be taken in order to correct the spurious numerical effects. 
\citet{CBH1996} devised to correct the finite volume effect to explore 
the scaling properties of the PDFs. Nevertheless, the general tendency 
between our simulations and those obtained by \citet{CBH1996} is quite similar,
i.e., the non-linear correction to the skewness and the kurtosis seen in the 
N-body results is generally small and their amplitudes are nearly 
constant on scales.

In figures \ref{fig:cumulants_scalefree_SCM} and
\ref{fig:cumulants_scalefree_ECM}, repeating the same procedure as in
the $\Lambda$CDM case, we plot the cumulant predictions from the PDF
taking account of the cutoff: {\it open-triangles} (cutoff-1) and {\it
filled-triangles} (cutoff-2). Then, the overall behaviors of the model
predictions seem to be greatly improved. The agreement between the
model prediction and the N-body simulation is reasonably good in both
cutoff-1 and cutoff-2 cases except for the strongly non-linear regime,
$\sigma_l\simgt5$ . A closer look at these figures shows that the
prediction based on the ECM provides a better approximation than that
of the SCM especially for the $n=0$ case. Note, however, that
most of the prediction for the variance slightly overpredict the
simulation except for the $n=-2$ case. The overprediction might be
partially ascribed to the cutoff of the density $[\delta_{\rm
min},\delta_{\rm max}]$, since the the width of the PDF quantified by
the variance sensitively depends on the cutoff.  On the other hand,
the reduced amplitudes $S_3$ and $S_4$ rather characterize the shape
of the PDF, which could be, in principle, less sensitive to the cutoff
of the high-density tails than cumulants themselves if the tails of PDF
converges to zero
enough. Actually, the shape of the predicted PDF resembles that 
of the PDF in the simulations as in figure \ref{fig:PDFscalefreel}.
The same tendency has been previously reported by \citet{FG1998}. 

Thus, at a level of the quality of N-body data, the local approximation with 
both the SCM and the ECM works reasonably well and the approximation with ECM 
even slightly improves the prediction. Of course, one must still care about 
the  cutoff density arising from the spurious numerical effects. In this 
respect, the validity of the local approximation may just reach at an 
acceptable level. In addition, another important caveat is drawn from   
the strong model dependence of the predictions.  
As discussed in section \ref{subsubsec:comparison}, the 
local approximation itself becomes more sensitive to the choice of the
Lagrangian local dynamics as the deviation from the spectral index
$n=-3$ gets larger.  This is clearly seen in the strong model
dependence of the one-point PDF (Fig. \ref{fig:PDFscalefreel})
and the non-linear correction for the cumulant prediction
(Fig. \ref{fig:ScaleFreeMoment}).  Further recall the fact that the
linear variance $\sigma_l$ scales as $\sigma_l\propto
R^{-(n+3)/2}$. This means that slight decrease of the smoothing radius
$R$ significantly increases the linear variance on non-linear scales,
$\sigma_l\simgt1$. Hence, one expects that the model prediction for
$n\geq0$ suffers from the non-linear corrections more seriously,
compared to the initial spectra $n=-2,-1$ or $\Lambda$CDM. Therefore,
the local approximation with ECM should be used with caution for the
index $n\geq0$.

%
%
%
%
%
%
%
%
%
%%%%%%%%%%%%%%%%%%%%%%%%%%%%%%%%%%%%%%%%%%%%%%%%%%%%%%%%%%%%%%%%%%%%%%
%%%%%%%%%%%%%%%%%%%%%%%%%%%%%%%%%%%%%%%%%%%%%%%%%%%%%%%%%%%%%%%%%%%%%%
\section{Conclusion and discussion}
\label{sec:conclusion}
%%%%%%%%%%%%%%%%%%%%%%%%%%%%%%%%%%%%%%%%%%%%%%%%%%%%%%%%%%%%%%%%%%%%%%
%%%%%%%%%%%%%%%%%%%%%%%%%%%%%%%%%%%%%%%%%%%%%%%%%%%%%%%%%%%%%%%%%%%%%%
%
%
%
%
%
%
%
%
In the present paper, we critically examined the validity and the
usefulness of the local approximation to the PDF and the cumulant
predictions. Adopting the ellipsoidal and the spherical collapse model
as representative model of the Lagrangian local dynamics, the PDFs and
the cumulants are calculated taking account of the smoothing
effect and the resultant predictions 
are compared with the N-body simulations with a Gaussian initial
condition. Due to the cutoff of the density arising from the 
spurious numerical effects, the detailed comparison in cumulants 
becomes difficult, and the correction for the cutoff density should be 
self-consistently incorporated into the model prediction. 
At a level of the quality of the N-body data, however, 
the local approximation with both SCM and ECM successfully
reproduces the N-body results for the PDFs and the cumulants, 
although a self-consistent calculation of local approximation 
presented in this paper (labeled by ``cutoff-2'') is still needed to be 
improved. This is indeed the case of the $\Lambda$CDM model and the 
scale-free models with indices $n=-2,~-1$.  
For the scale-free model with $n=0$, while the discrepancy between 
the model prediction and the simulation result is manifest in the local 
approximation with SCM,  
the agreement with N-body results still remains good for the ECM prediction. 
The detailed discussion reveals that the prediction based on the local 
approximation sensitively depends on the slope of the initial spectrum 
and that the predictions for $n>0$ become more sensitive to the non-linear 
dynamics of the local collapse model. 
Thus, a more delicate modeling of the Lagrangian 
local dynamics is required for an accurate prediction. 
Taking this point carefully, we therefore conclude that 
the local approximation with SCM and ECM 
provides an excellent approximation to the N-body simulations 
for CDM and scale-free models with $n<0$ 
in both the linear and the non-linear regimes, $0\simlt \sigma_l\simlt 5$, 
while the local approximation should be used with caution in 
the $n\geq0$ cases.

In this paper, we found that the predictions based on the ellipsoidal
collapse model somewhat improve the approximation, however, the degree
of the improvement is not so large as long as a CDM-like initial
spectrum (i.e., effective spectral index
$n_{\rm eff}=-3-d\log\sigma_l^2/d\log R<0$)
is concerned. Compared to the prediction from the spherical collapse model,
the calculation of PDF from ellipsoidal collapse model is rather
complicated and require a time-consuming numerical integration.
It seems that the spherical collapse model provides a simpler
prescription for the PDF in real space and is practically more useful
than the ellipsoidal collapse model. However, if one considers the
one-point statistics in redshift space, the situation might be changed
drastically. As reported by \citet{SG2001}, the local approximation with
spherical collapse model only provides a good approximation to the
redshift space PDFs when $\sigma_l\simlt0.4$. A part of this reason is
ascribed to the fact that the model prediction cannot recover the
linear perturbation result, referred to as the Kaiser effect \citep{K1987}; 
the variance in the redshift space, $\sigma_z^2$ is related to the one in
the real space as:
%%%%%%%%%%%%%%%%%%%%%%%%%%%%%%%%%%%%%%%%%%%%%%%%%%%%%%%%%%%%%%%%%%%%%%
\begin{equation}
 \sigma_z^2=\left(1+\frac{2}{3}f_\Omega+\frac{1}{5}f_\Omega^2\right)\sigma_l^2.
\label{eq:Kaiser_effect}
\end{equation}
%%%%%%%%%%%%%%%%%%%%%%%%%%%%%%%%%%%%%%%%%%%%%%%%%%%%%%%%%%%%%%%%%%%%%%
In contrast to the spherical collapse model, which leads to the 
incorrect prediction $\sigma_z^2=(1+f_\Omega/3)^2\sigma_l^2$,  
the Kaiser effect (\ref{eq:Kaiser_effect}) can be correctly recovered by means of the
ellipsoidal collapse model. The derivation of equation
(\ref{eq:Kaiser_effect}) is presented in appendix
\ref{appen:Kaiser_ECM}. This fact is very interesting and also provides
an important suggestion that the non-sphericity of the Lagrangian local
dynamics play a crucial role in computing the one-point statistics in
redshift space and is indeed essential for an accurate prediction. The
detailed analysis of the model predictions in redshift space is now in
progress and will be described elsewhere.
%
%
%
%
%
%
%
%%%%%%%%%%%%%%%%%%%%%%%%%%%%%%%%%%%%%%%%%%%%%%%%%%%%%%%%%%%%%%%%%%%%%%
%%%%%%%%%%%%%%%%%%%%%%%%%%%%%%%%%%%%%%%%%%%%%%%%%%%%%%%%%%%%%%%%%%%%%%
\acknowledgments
%%%%%%%%%%%%%%%%%%%%%%%%%%%%%%%%%%%%%%%%%%%%%%%%%%%%%%%%%%%%%%%%%%%%%%
%%%%%%%%%%%%%%%%%%%%%%%%%%%%%%%%%%%%%%%%%%%%%%%%%%%%%%%%%%%%%%%%%%%%%%
We thank the referee, E. Gazta\~naga for many useful comments and suggestions  
to improve the original manuscript. 
We also thank Y. P. Jing for kindly providing us his N-body data and Y. Suto
for comments and discussions. 
I.K. acknowledges support from a Takenaka-Ikueikai Fellowship. This work
is supported in part by a Grant-in-Aid for Scientific Research from the
JSPS(No.$14740157$).
\clearpage
\appendix
%%%%%%%%%%%%%%%%%%%%%%%%%%%%%%%%%%%%%%%%%%%%%%%%%%%%%%%%%%%%%%%%%%%%%%
%%%%%%%%%%%%%%%%%%%%%%%%%%%%%%%%%%%%%%%%%%%%%%%%%%%%%%%%%%%%%%%%%%%%%%
\section{Derivation of Kaiser factor by ellipsoidal collapse model}
\label{appen:Kaiser_ECM}
%%%%%%%%%%%%%%%%%%%%%%%%%%%%%%%%%%%%%%%%%%%%%%%%%%%%%%%%%%%%%%%%%%%%%%
%%%%%%%%%%%%%%%%%%%%%%%%%%%%%%%%%%%%%%%%%%%%%%%%%%%%%%%%%%%%%%%%%%%%%%
%
%
%
%
%
%
%
%
%
%
In this appendix, we derive the Kaiser effect (\ref{eq:Kaiser_effect})
from the ellipsoidal collapse model. Assuming the distant-observer
approximation, let us consider the local density located at
$\bm{r}=(r_1,r_2,r_3)$ in real space and choose the third axis as the
line-of-sight direction. Denoting the corresponding coordinate in the
redshift space by $\bm{s}=(s_1,s_2,s_3)$, the relation $\bm{r}$ and
$\bm{s}$ becomes
%%%%%%%%%%%%%%%%%%%%%%%%%%%%%%%%%%%%%%%%%%%%%%%%%%%%%%%%%%%%%%%%%%%%%%%%%%%%
\begin{equation}
 s_1=r_1,~~s_2=r_2,~~s_3=r_3+v_3/H,
\end{equation}
%%%%%%%%%%%%%%%%%%%%%%%%%%%%%%%%%%%%%%%%%%%%%%%%%%%%%%%%%%%%%%%%%%%%%%%%%%%%
where $v_3$ is the line-of-sight component of the peculiar velocity
field. Then the local density in redshift space, $\rho_s$  can be
expressed in terms of the quantities in real space as follows:
%%%%%%%%%%%%%%%%%%%%%%%%%%%%%%%%%%%%%%%%%%%%%%%%%%%%%%%%%%%%%%%%%%%%%%%%%%%%
\begin{equation}
 \rho_s = \frac{dM}{ds_1ds_2ds_3}=
  \frac{dM}{\displaystyle dr_1dr_2dr_3
  \left(1+\frac{1}{H}\frac{\partial v_3}{\partial r_3}\right)}
  =\frac{\rho}{\displaystyle 1+\frac{1}{H}\frac{\partial v_3}{\partial r_3}}.
  \label{eq:rho_s_rho_r}
\end{equation}
%%%%%%%%%%%%%%%%%%%%%%%%%%%%%%%%%%%%%%%%%%%%%%%%%%%%%%%%%%%%%%%%%%%%%%%%%%%%
The peculiar velocity field at the position $\bm{r}$ is described by the
motion of the homogeneous ellipsoid. Introducing the new coordinate
along the principal axis of ellipsoid, $\bm{r}'=(r'_1,r'_2,r'_3)$, the
peculiar velocity $\bm{v}$ is given by
%%%%%%%%%%%%%%%%%%%%%%%%%%%%%%%%%%%%%%%%%%%%%%%%%%%%%%%%%%%%%%%%%%%%%%%%%%%%
\begin{equation}
 v_i =\bar{v}_i+ \left(\frac{\dot{\alpha}_i}{\alpha_i}-H\right)\,r'_i
  \label{appen:v_ellip}
\end{equation}
%%%%%%%%%%%%%%%%%%%%%%%%%%%%%%%%%%%%%%%%%%%%%%%%%%%%%%%%%%%%%%%%%%%%%%%%%%%%
in the new coordinate system. Here, $\bar{v}_i$ is the bulk velocity of
the ellipsoid. The new coordinate $\bm{r}'$ does not necessarily
coincides with the original one $\bm{r}$. Rather, it is related to the
original coordinate through the Euler angle, i.e.,
$\bm{r}'= R_3(\psi)R_2(\theta)R_3(\phi)\bm{r}$, where $R_i$ is the
rotational matrix with respect to the $i$-axis. Using this fact, the
quantity (\ref{appen:v_ellip}) is transformed into the original frame
and we obtain
%%%%%%%%%%%%%%%%%%%%%%%%%%%%%%%%%%%%%%%%%%%%%%%%%%%%%%%%%%%%%%%%%%%%%%%%%%%%
\begin{eqnarray}
 \frac{\partial v_3}{\partial r_3}&=& \sum_{i=1}^3\,\,
  \left[R_3(\psi)R_2(\theta)R_3(\phi)\right]^{-1}_{3i}\,\,
  \left(\frac{\dot{\alpha_i}}{\alpha_i}-H\right)\,\,
  \left[R_3(\psi)R_2(\theta)R_3(\phi)\right]_{i3} \nonumber \\
 &=&\left(\frac{\dot{\alpha_1}}{\alpha_1}-H\right)\cos^2\psi\sin^2\theta+
  \left(\frac{\dot{\alpha_2}}{\alpha_2}-H\right)\sin^2\psi\sin^2\theta+
  \left(\frac{\dot{\alpha_3}}{\alpha_3}-H\right)\cos^2\theta.
\end{eqnarray}
%%%%%%%%%%%%%%%%%%%%%%%%%%%%%%%%%%%%%%%%%%%%%%%%%%%%%%%%%%%%%%%%%%%%%%%%%%%%
Here, we neglect the bulk velocity. Substituting the above equation into
(\ref{eq:rho_s_rho_r}) yields
%%%%%%%%%%%%%%%%%%%%%%%%%%%%%%%%%%%%%%%%%%%%%%%%%%%%%%%%%%%%%%%%%%%%%%
\begin{equation}
 1+\delta_s=\frac{1+\delta}{\displaystyle \frac{1}{H}\left(
   \frac{\dot{\alpha_1}}{\alpha_1}\cos^2\psi\sin^2\theta
   +\frac{\dot{\alpha_2}}{\alpha_2}\sin^2\psi\sin^2\theta
   +\frac{\dot{\alpha_3}}{\alpha_3}\cos^2\theta\right)},
  \label{eq:del_s_del_r} 
\end{equation}
%%%%%%%%%%%%%%%%%%%%%%%%%%%%%%%%%%%%%%%%%%%%%%%%%%%%%%%%%%%%%%%%%%%%%%
with the quantity $\delta_s$ being the density fluctuation in redshift
space. Note that the expression is exact under the distant-observer
limit. In the linear perturbation, the quantity $\alpha_i$ in
(\ref{eq:del_s_del_r}) is replaced with $a(1-\lambda_i)$
(eq.[\ref{eq:init_alpha}]). We have
%%%%%%%%%%%%%%%%%%%%%%%%%%%%%%%%%%%%%%%%%%%%%%%%%%%%%%%%%%%%%%%%%%%%%%
\begin{equation}
 \delta_s=\delta_l+f_\Omega\,\,(\lambda_1\cos^2\psi\sin^2\theta+
  \lambda_2\sin^2\psi\sin^2\theta+\lambda_3\cos^2\theta), 
\end{equation}
%%%%%%%%%%%%%%%%%%%%%%%%%%%%%%%%%%%%%%%%%%%%%%%%%%%%%%%%%%%%%%%%%%%%%%
where $f_\Omega\equiv d\ln D/d\ln a\simeq\Omega_{\rm m}^{0.6}$. Hence,
the linear variance in redshift space, $\sigma_z^2$ is expressed as
%%%%%%%%%%%%%%%%%%%%%%%%%%%%%%%%%%%%%%%%%%%%%%%%%%%%%%%%%%%%%%%%%%%%%%
\begin{equation}
 \sigma_z^2\equiv \langle\delta_s^2\rangle=
  \left(1+\frac{2}{3}f_\Omega\right)
  \langle(\lambda_1+\lambda_2+\lambda_3)^2\rangle
  +\frac{1}{15}f_\Omega^2\langle 3(\lambda_1^2+\lambda_2^2+\lambda_3^2)+
  2(\lambda_1\lambda_2+\lambda_2\lambda_3+\lambda_3\lambda_1) \rangle, 
\label{appen:sigma_z}
\end{equation}
%%%%%%%%%%%%%%%%%%%%%%%%%%%%%%%%%%%%%%%%%%%%%%%%%%%%%%%%%%%%%%%%%%%%%%
where we have taken the averages over the angles, $\psi$, $\theta$ and
$\phi$. Finally, the ensemble averages over the variable $\lambda_i$ are
taken with a knowledge of the distribution function (\ref{eq:DoroPDF}):
%%%%%%%%%%%%%%%%%%%%%%%%%%%%%%%%%%%%%%%%%%%%%%%%%%%%%%%%%%%%%%%%%%%%%%
\begin{equation}
 \sigma_z^2=\left(1+\frac{2}{3}f_\Omega+\frac{1}{5}f_\Omega^2\right)\sigma_l^2.
\end{equation}
%%%%%%%%%%%%%%%%%%%%%%%%%%%%%%%%%%%%%%%%%%%%%%%%%%%%%%%%%%%%%%%%%%%%%%
This is exactly the Kaiser effect. Note that in the cases adopting the spherical 
collapse model, the variable $\lambda_i$ means $\delta_l/3$. Hence, the 
ensemble average over $\delta_l$ in equation (\ref{appen:sigma_z}) immediately yields 
incorrect prediction, $\sigma_z^2=(1+f_\Omega/3)^2\sigma_l^2$.

\clearpage
%%%%%%%%%%%%%%%%%%%%%%%%%%%%%%%%%%%%%%%%%%%%%%%%%%%%%%%%%%%%%%%%%%%%%%
%%%%%%%%%%%%%%%%%%%%%%%%%%    REFERENCES    %%%%%%%%%%%%%%%%%%%%%%%%%%
%%%%%%%%%%%%%%%%%%%%%%%%%%%%%%%%%%%%%%%%%%%%%%%%%%%%%%%%%%%%%%%%%%%%%%

%%%%%%%%%%%%%%%%%%%%%%%%%%%%%%%%%%%%%%%%%%%%%%%%%%%%%%%%%%%%%%%%%%%%%%
%
%
%
%
%
%
%
%
%
%
%
%
%
%
%
%
%
%
%
\clearpage
%%%%%%%%%%%%%%%%%%%%%%%%%%%%%%%%%%%%%%%%%%%%%%%%%%%%%%%%%%%%%%%%%%%%%%
%%%%%%%%%%%%%%%%%%%%%%%%%%%    TABLES     %%%%%%%%%%%%%%%%%%%%%%%%%%%%
%%%%%%%%%%%%%%%%%%%%%%%%%%%%%%%%%%%%%%%%%%%%%%%%%%%%%%%%%%%%%%%%%%%%%%
\begin{deluxetable}{c|cccc|cccc}
 \tablecaption{Coefficients of perturbative correction for the cumulants
 of the density field
 \label{tbl:coeff}}
 \tablehead{
 & \multicolumn{4}{c|}{non-linear external tide} &
 \multicolumn{4}{|c}{linear external tide}\\
 $n$ & -3 &-2 &-1 & 0 & -3 &-2 &-1 & 0
 }
 \startdata
 $s_{2,4}$ & 1.79 & 0.772 & 0.363 & 0.566 & 1.73 & 0.708 & 0.299 & 0.501\\
 $s_{2,6}$ & 4.52 & 0.539 & 0.0364 & 0.360 & 4.20 & 0.442 & $-0.0202$ & 0.165\\
 $s_{2,8}$ & 14.8 & 0.294 & $-7.50\times 10^{-4}$ & 0.158 & 13.3 & 0.199
 & $-0.0113$ & $-0.235$\\ \hline
 $S_{3,2}$ & 10.2 & 3.19 & 0.525 & 0.00379 & 10.2 & 3.19 & 0.587 & 0.130\\
 $S_{3,4}$ & 53.2 & 4.20 & $-0.0169$ & 0.0646 & 51.7 & 3.98 & 0.0119 & 0.246\\ \hline
 $S_{4,2}$ & 271 & 63.2 & 6.67 & $-0.0227$ & 269 & 63.1 & 7.32 & 0.651\\
 $S_{4,4}$ & 2.19$\times 10^3$ & 146 & 0.614 & 0.364 & 2.14$\times 10^3$
 & 141 & 1.16 & 1.41\\ \hline
 \enddata
\end{deluxetable}
%%%%%%%%%%%%%%%%%%%%%%%%%%%%%%%%%%%%%%%%%%%%%%%%%%%%%%%%%%%%%%%%%%%%%%
%%%%%%%%%%%%%%%%%%%%%%%%%%%    FIGURES    %%%%%%%%%%%%%%%%%%%%%%%%%%%%
%%%%%%%%%%%%%%%%%%%%%%%%%%%%%%%%%%%%%%%%%%%%%%%%%%%%%%%%%%%%%%%%%%%%%%
%%%%%%%%%%%%%%%%%%%%%%%%%%%%%%%%%%%%%%%%%%%%%%%%%%%%%%%%%%%%%%%%%%%%%%
%%%%%%%%%%%%%%%%%%%%%%%%%%%%%%%%%%%%%%%%%%%%%%%%%%%%%%%%%%%%%%%%%%%%%%
%%%%%%%%%%%%%%%%%%%%%%%%%%%%%%%%%%%%%%%%%%%%%%%%%%%%%%%%%%%%%%%%%%%%%%
%%%%%%%%%%%%%%%%%%%%%%%%%%%%%%%%%%%%%%%%%%%%%%%%%%%%%%%%%%%%%%%%%%%%%%
%%%%%%%%%%%%%%%%%%%%%%%%%%%%%%%%%%%%%%%%%%%%%%%%%%%%%%%%%%%%%%%%%%%%%%
%%%%%%%%%%%%%%%%%%%%%%%%%%%%%%%%%%%%%%%%%%%%%%%%%%%%%%%%%%%%%%%%%%%%%%
%%%%%%%%%%%%%%%%%%%%%%%%%%%%%%%%%%%%%%%%%%%%%%%%%%%%%%%%%%%%%%%%%%%%%%
%%%%%%%%%%%%%%%%%%%%%%%%%%%%%%%%%%%%%%%%%%%%%%%%%%%%%%%%%%%%%%%%%%%%%%

\begin{thebibliography}{}
 \bibitem[Bernardeau(1992)]{B1992}
   Bernardeau, F. 1992, \apj, 392, 1
 \bibitem[Bernardeau(1994a)]{B1994a}
   Bernardeau, F. 1994a, \aap, 291, 697
 \bibitem[Bernardeau(1994b)]{B1994b}
   Bernardeau, F. 1994b, \apj 433, 1
 \bibitem[Bernardeau \& Kofman(1995)]{BK1995}
   Bernardeau, F., \& Kofman, L. 1995, \apj, 443, 479
 \bibitem[Bond \& Myers(1996)]{BM1996}
   Bond, J. R., \& Myers, S.T. 1996, \apjs, 103, 1
 \bibitem[Coles \& Jones(1991)]{CJ1991}
   Coles, P., \& Jones, B. 1991, \mnras 248, 1
 \bibitem[Coles, Melott, \& Shandarin(1993)]{CMS1993}
   Coles, P., Melott, A. L., \& Shandarin, S. F. 1993, \mnras, 260, 765
 \bibitem[Colombi, Bouchet \& Hernquist(1996)]{CBH1996}
   Colombi, S., Bouchet, F. R., \& Hernquist, L. 1996, \apj, 465, 14 	 
 \bibitem[Doroshkevich(1970)]{D1970}
   Doroshkevich, A. G. 1970, Astrofizika, 6, 581
 \bibitem[Fosalba \& Gazta\~naga(1998a)]{FG1998}
   Fosalba, P., \& Gazta\~naga, E. 1998a, \mnras, 301, 503
 \bibitem[Fosalba \& Gazta\~naga(1998b)]{FG1998b}
   Fosalba, P., \& Gazta\~naga, E. 1998b, \mnras, 301, 535
 \bibitem[Gazta\~naga \& Yokoyama(1993)]{GY1993}
   Gazta\~naga, E., \& Yokoyama, J. 1993, \apj, 403, 450
 \bibitem[Hamilton(1985)]{H1985}
   Hamilton, A. J. S. 1985, \apj, 292, L35
 \bibitem[Jing(1998)]{J1998}
   Jing, Y. P. 1998, \apj, 503, L9
 \bibitem[Jing \& Suto(1998)]{JS1998}
   Jing, Y. P., \& Suto, Y. 1998, \apj, 494, L5
 \bibitem[Kaiser(1987)]{K1987} 
   Kaiser, N. 1987, \mnras, 227, 1 	
 \bibitem[Kayo, Taruya \& Suto(2001)]{KTS2001}
   Kayo, I., Taruya, A., \& Suto, Y. 2001, \apj, 561, 22
 \bibitem[Nakamura \& Suto(1995)]{NS1995}
    Nakamura, T. T., \& Suto, Y. 1995, \apj, 447, 65
 \bibitem[Ohta, Kayo \& Taruya(2003)]{OKT2003}
   Ohta, Y., Kayo, I., \& Taruya, A. 2003 \apj, 589, 1
 \bibitem[Saslaw \& Hamilton(1984)]{SH1984}
   Saslaw, W. C., \& Hamilton, A. J. S. 1984 \apj, 276, 13
 \bibitem[Scherrer \& Gazta\~naga(2001)]{SG2001} 
   Scherrer, R. J., \& Gazta\~naga, E. 2001, \mnras, 328, 257
 \bibitem[Scoccimarro(1997)]{S1997}
   Scoccimarro, R. 1997, \apj, 487, 1
 \bibitem[Scoccimarro \& Frieman(1996)]{SF1996b}
   Scoccimarro, R., \& Frieman, J. A. 1996, \apj, 473, 620
 \bibitem[Taruya, Hamana \& Kayo(2003)]{THK2003} 
   Taruya, A., Hamana, T., \& Kayo, I. 2003, \apj, 339, 495	
 \bibitem[Taylor \& Watts(2000)]{TW2000}
   Taylor, A. N., \& Watts, P. I. R., 2000, \mnras, 314, 92
 \bibitem[Ueda \& Yokoyama(1996)]{UY1996}
   Ueda, H., \& Yokoyama, J. 1996, \mnras  280, 754
\end{thebibliography}
\end{document}